\documentclass[aps,prl,twocolumn,footinbib,superscriptaddress]{revtex4-1} 

	\usepackage{textcomp}
	\usepackage{amsmath}
	\usepackage{amssymb}
	\usepackage{graphicx} 
	\usepackage[breaklinks=true,colorlinks,citecolor=blue,linkcolor=blue,urlcolor=blue]{hyperref}
	\usepackage{mathtools}
	\usepackage{bm}
	\usepackage{color}
	\usepackage{textcomp}
	\usepackage{ulem}
	\usepackage{braket}

	\renewcommand{\i}{i}
	
	\renewcommand{\Re}{\mathop{\text{Re}}\nolimits}

\begin{document}
\title{Non-Linear Thermovoltage in a Single-Electron Transistor}

\author{P. A. Erdman}
\affiliation{NEST, Scuola Normale Superiore and Istituto Nanoscienze-CNR, I-56127 Pisa, Italy}
\email{paolo.erdman@sns.it}

\author{J. T. Peltonen}
\affiliation{QTF Centre of Excellence, Department of Applied Physics, Aalto University School of Science, P.O. Box 13500, 00076 Aalto, Finland}
	
\author{B. Bhandari}
\affiliation{NEST, Scuola Normale Superiore and Istituto Nanoscienze-CNR, I-56127 Pisa, Italy}
\email{bibek.bhandari@sns.it}

\author{B. Dutta}
\affiliation{Univ. Grenoble Alpes, CNRS, Institut N\'eel, 25 Avenue des Martyrs, 38042 Grenoble, France}

\author{H. Courtois}
\affiliation{Univ. Grenoble Alpes, CNRS, Institut N\'eel, 25 Avenue des Martyrs, 38042 Grenoble, France}

\author{R. Fazio}
\affiliation{ICTP, Strada Costiera 11, I-34151 Trieste, Italy}
\affiliation{NEST, Scuola Normale Superiore and Istituto Nanoscienze-CNR, I-56127 Pisa, Italy}

\author{F. Taddei}
\affiliation{NEST, Istituto Nanoscienze-CNR and Scuola Normale Superiore, I-56126 Pisa, Italy}

\author{J. P. Pekola}
\affiliation{QTF Centre of Excellence, Department of Applied Physics, Aalto University School of Science, P.O. Box 13500, 00076 Aalto, Finland}

\begin{abstract}
We perform direct thermovoltage measurements in a single-electron transistor, using on-chip local thermometers, both in the linear and non-linear regimes. Using a model which accounts for co-tunneling, we find excellent agreement with the experimental data with no free parameters even when the temperature difference is larger than the average temperature (far-from-linear regime). 
This allows us to confirm the sensitivity of the thermovoltage on co-tunneling and to find that in the non-linear regime the temperature of the metallic island is a crucial parameter. Surprisingly, the metallic island tends to overheat even at zero net charge current, resulting in a reduction of the thermovoltage. 
\end{abstract}



\maketitle
{\it Introduction.}---The use of nano-devices has emerged as one of the key technologies in the quest to establish a sustainable energy system, allowing at the same time the control of heat flow in small circuits \cite{benenti2017}.
So far, most of the investigations of thermal properties in nanostructures have focused on the thermal conductance \cite{molenkamp1992,schwab2000,chiatti2006,meschke2006,hoffmann2009,jezouin2013,banerjee2017,dutta2017,cui2017,mosso2017}. 
Conversely the thermovoltage, which describes the electrical response to a temperature difference  and is directly related to both the power and efficiency of thermal machines \cite{benenti2017}, is much less studied. This is due to the difficulty in coupling local sensitive electron thermometers and heaters/coolers to the sample under study in order to have a well-defined, known temperature difference across the device.
The thermovoltage has been measured in devices based on nanowires \cite{roddaro2013,wu2013} and on quantum dots \cite{dzurak1993, staring1993,molenkamp1994, dzurak1997, moller1998, godijn1999, llaguno2004, scheibner2005, pogosov2006, scheibner2007, svensson2012, svensson2013, dutta2018}. In these experiments, however,
 the temperature of the electrodes were typically not measured directly, but rather determined as fitting parameters, and there are no experiments where the temperature of the electrodes and the thermovoltage are measured simultaneously.
Furthermore, there are no experiments probing the thermovoltage in devices based on metallic islands, while theoretical works for these systems have focused only on the linear response regime \cite{beenakker1992,andreev2001,turek2002,koch2004, kubala2006,kubala2008,vasenko2015,nonlin_note}.

In this paper, we report for the first time on the measurement of the thermovoltage in a metallic single-electron transistor (SET) using on-chip, local tunnel-junction-based thermometers and electron temperature control. This system allows us to perform thermoelectric measurements with an unprecedented control, both within the linear and non-linear response regimes, imposing temperature differences exceeding the average temperature.
Using a theoretical model which accounts for non-linear effects and co-tunneling processes, we find an excellent agreement with the experimental data with no free parameters. On one hand, this allows us to nail down quantitatively the role of co-tunneling processes on the thermovoltage. 
On the other hand, we find that in the non-linear regime the temperature of the island emerges as a crucial parameter. Surprisingly, although the thermovoltage is measured at zero net charge current, within the non-linear response the island tends to overheat to a temperature greater than the average lead temperature, which results in a suppression of the thermovoltage. We show, however, that the non-linear thermovoltage can be optimized up to a factor two with respect to the experimentally observed value by lowering the temperature of the island to the temperature of the cold lead. This could be achieved by exploiting the phonons in the island which act as a third thermal bath coupled to our system.

\begin{figure}[!t]
	\centering
	\includegraphics[width=1\columnwidth]{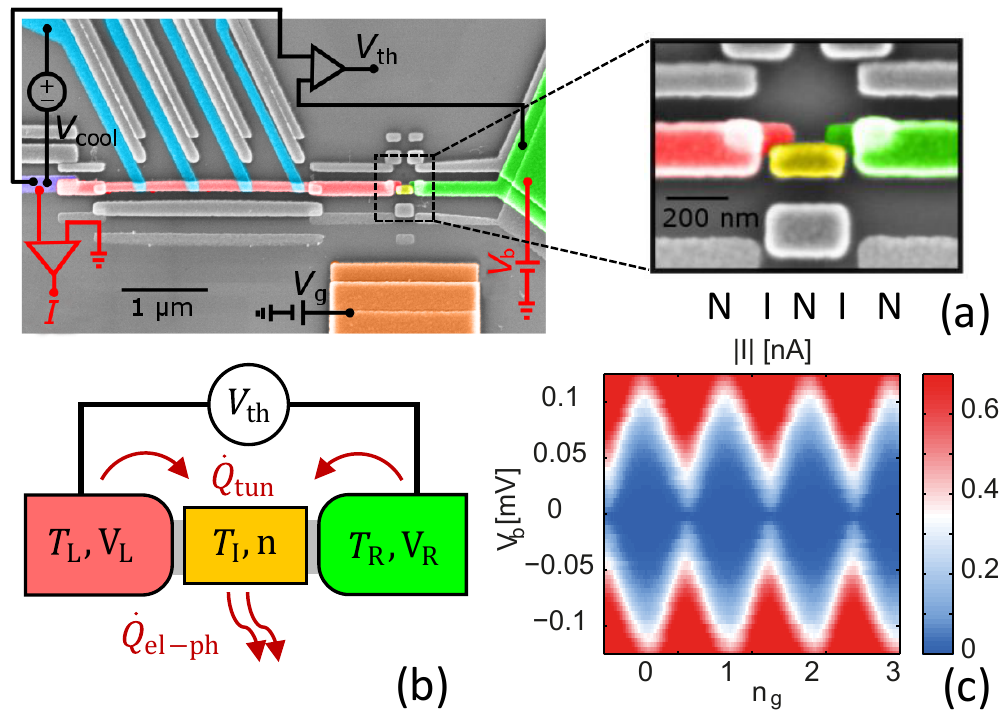}
	\caption{Representation and characterization of the single-electron transistor. a) False-colored SEM image of the full device and a zoomed in view around the metallic island (yellow) tunnel coupled to two normal leads (red and green). b) Schematic representation of the system with the same coloring as in the SEM image. The heat balance in the metallic island is represented by red arrows. c) Absolute value of the current through the SET as a function of the applied source drain voltage $V_\text{b}$ and of the gate-induced charge $n_g$.}
	\label{fig:setup}
\end{figure}
{\it The experimental setup.}---
Fig.~\ref{fig:setup}a) is a colored scanning electron micrograph of the device and Fig.~\ref{fig:setup}b) is a schematic representation 
of the experiment with the same colors highlighting the main elements of the fully normal-conducting SET. 
The left lead L (red) and right lead R (green) are tunnel and capacitively coupled to a central metallic island I (yellow), which is under the influence of a tunable gate electric field (orange). A voltage bias $V_\text{b}=V_\text{L}-V_\text{R}$ can be applied to the SET electrodes and the corresponding current $I$ can be measured for an initial characterization of the device. The temperature $T_\text{R}$ of the electrons in R is fixed to the bath temperature, given the strong electron-phonon coupling in the large and ``bulky'' lead. On the other hand, the electronic temperature $T_\text{L}$ in the left lead (red) can both be varied and measured using the superconducting tunnel probes (blue). The tunability of the temperature is possible thanks to the superconducting wire (purple) in clean contact with the left lead through which there is no heat conduction, and thanks to the limited size of the normal (red) part of the lead that reduces the electron-phonon heat flux. Electrons within the island are in local equilibrium at temperature $T_\text{I}$ since the electron-electron interaction is much faster than the tunneling rates \cite{giazotto2006}. 
The experiment is performed in a dilution refrigerator at bath temperatures typically between 50 and 400 mK. For the thermovoltage measurements, the SET voltage bias source and current preamplifier (sketched in red in Fig. 1a) are disconnected. Crucially, the thermovoltage $V_{\text{th}}$ is probed directly across the SET using a room-temperature voltage preamplifier with ultralow input bias current below 20 fA. 
Fabrication details can be found in Ref.~\cite{dutta2017} where ``sample B'' is the device used for this experiment.

Figure~\ref{fig:setup}c) shows the absolute value of the current $I$ across the device at $65 \ \mathrm{mK}$ as a function of the potential bias $V_\text{b}$ and of the gate-induced charge $n_g= (C_\text{L}V_\text{L} + C_\text{R}V_\text{R} + C_\text{g}V_\text{g})/e$ , where $C_\text{L}$, $C_\text{R}$ and $C_\text{g}$ are, respectively, the capacitances of the island to L, R and to the gate electrode, and $V_\text{g}$ is the gate voltage. In the dark blue regions, Coulomb diamonds, single electron tunneling between the leads and the island is not allowed, and the current is very small. At half integer values of $n_g$, ``degeneracy points'', there are conductance peaks at zero bias since single electron tunneling is allowed for any finite voltage bias.

{\it The model.}---The state of the SET is characterized by the probability $P(n)$ to have $n$ excess charges on the island. The electrostatic energy necessary for this is
\begin{equation}
	U(n) = E_C \left( n-n_g \right)^2,
\end{equation}
where $E_C = e^2/(2C)$ is the charging energy with $C=C_\text{L}+C_\text{R} +C_\text{g}$. Electron tunneling between the leads and the island induces transitions between charge states. 
The leading order process in a perturbative expansion in the tunnel coupling between the island and the leads corresponds to a single electron transfer between the leads and the island (sequential tunnelling) \cite{averin1991,nazarov2009}. The sequential-tunneling rates for transferring electrons from $\alpha=\text{L,R (I)}$ to $\beta= \text{I (L,R)}$, with the island initially having $n$ charges, is denoted by $\Gamma_{\alpha\beta}(n)$  (see Supplemental Material for details \cite{suppl}). 

Higher order processes can become dominant if all sequential-tunneling processes are energetically unfavorable [in the Coulomb diamond region in Fig.~\ref{fig:setup}c)]. In particular, co-tunneling (second order process) refers to the transfer of an electron from one lead to another, without changing the charge state of the island but going through a virtual state. 
The dominant contribution of this kind is inelastic co-tunneling, i.e. the electron which tunnels from lead L, say, to I via a virtual state has a different energy with respect to the electron tunneling from I to R
\cite{Note1}.
We denote the rate of inelastic co-tunneling that transfers a charge from $\alpha =\text{L (R)}$ to $\beta=\text{R (L)}$, when $n$ electrons are on the island before the process occurs, by $\gamma_{\alpha\beta}(n)$. 

The probabilities $P(n)$ can be computed by solving a master equation (see Supplemental Material for details \cite{suppl}).
The charge current can then be written as $I(V_\text{b}) = I^{\text{seq}} + I^{\text{cot}}$, where
\begin{equation}
	I^{\text{seq}} = e \sum_n P(n)\left[\Gamma_{LI}(n) - \Gamma_{IL}(n)  \right]
	\label{eq:i_seq}
\end{equation}
is the sequential-tunneling contribution, given by electrons tunneling between lead L and I, and
\begin{equation}
	I^{\text{cot}} = e \sum_n P(n)\left[\gamma_{LR}(n) - \gamma_{RL}(n)  \right]
	\label{eq:i_cot}
\end{equation}
is the inelastic co-tunneling contribution \cite{turek2002,koch2004,nazarov2009,kaasbjerg2016,bhandari2018}. We compute the sequential and co-tunneling rates exactly, without linearizing in the voltage bias and temperature difference (see Supplemental Material for details \cite{suppl}).

In the presence of a fixed temperature bias ($T_\text{R}\neq T_\text{L}$), the thermovoltage $V_{\text{th}}$ is the solution to
\begin{equation}
	I(V_{\text{th}})=0.
	\label{eq:vth_def}
\end{equation}
Notice that the charge current also depends on the temperature of the island $T_\text{I}$.
By imposing that the charge current and the net energy entering the island through electron tunneling are zero, we find that
\begin{equation}
 	T_\text{I} = \frac{T_\text{L}R_\text{R}+T_\text{R}R_\text{L}}{R_\text{L} + R_\text{R}},
 	\label{eq:ti_lin}
\end{equation}
where $R_\text{L}$ and $R_\text{R}$ are respectively the resistance of the left and right tunnel junctions.
Eq.~(\ref{eq:ti_lin}), which is found performing a simple sequential tunneling calculation within linear response and in the two charge state approximation (valid for $E_C \gg k_BT$), reduces to $T_{\text{I}} =\bar{T} \equiv (T_\text{L} + T_\text{R})/2$ in the present symmetric case where $R_\text{L}=R_\text{R}$. We will thus initially assume that $T_{\text{I}}$ is given by the average lead temperature $\bar{T}$. However, as we will soon discuss in detail, we find that this assumption gives quantitatively wrong results beyond the linear response regime, leading us to the exploration of the impact of $T_\text{I}$ on the thermovoltage.

\begin{figure}[!tb]
	\centering
	\includegraphics[width=1\columnwidth]{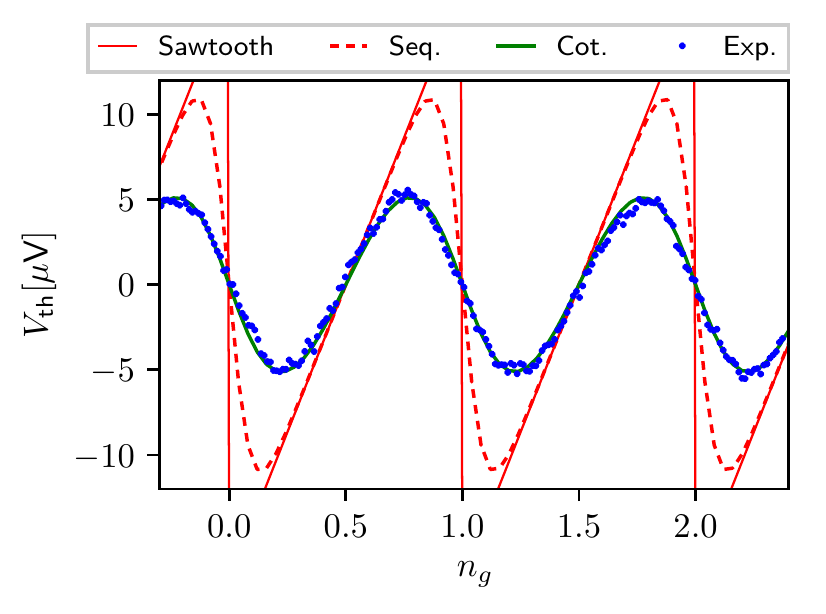}
	\caption{Experimental and theoretical thermovoltage as a function of $n_g$. The red thin curve represents the sawtooth behavior predicted with a sequential-tunneling calculation in linear response and accounting for two charge states. The dashed red curve is found by solving Eq.~(\ref{eq:vth_def}) including only sequential contributions, while the green curve includes also co-tunneling contributions. The temperatures of the leads are $T_\text{L}=134 \ \mathrm{mK}$ and $T_\text{R} = 190 \ \mathrm{mK}$ and, according to Eq.~(\ref{eq:ti_lin}), we assume that $T_\text{I}=\bar{T}$.}
	\label{fig:seq_compare}
\end{figure}

\begin{figure}[!tb]
	\centering
	\includegraphics[width=1\columnwidth]{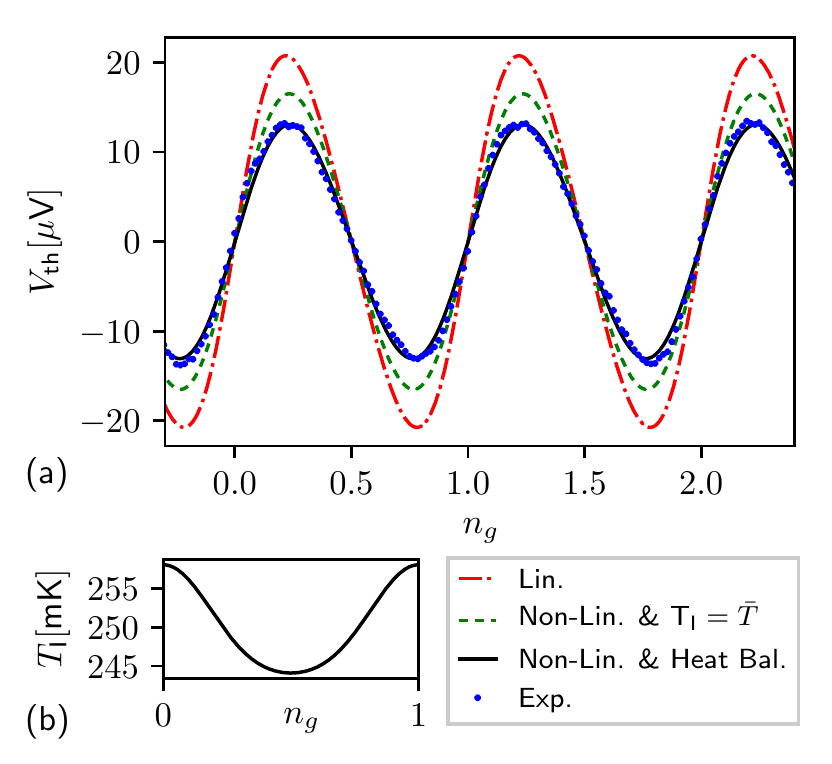}
	\caption{a) Experimental and theoretical thermovoltage as a function of $n_g$. All theoretical curves include co-tunneling. The red dashed-dotted curve corresponds to a linear response calculation around $\bar{T}$. The green dashed curve corresponds to a non-linear calculation where we fix $T_\text{I} = \bar{T}$, while the black curve corresponds to a non-linear calculation where $T_\text{I}$, shown in b) as a function of $n_g$, is calculated solving the heat balance condition in Eq.~(\ref{eq:heat_bal}) together with Eq.~(\ref{eq:vth_def}).  The temperatures of the leads are $T_\text{L}=342 \ \mathrm{mK}$ and $T_\text{R} = 63 \ \mathrm{mK}$.}
	\label{fig:nonlin_compare}
\end{figure}

{\it Results.}---We focus on two data sets which represent two different regimes: linear response (Fig.~\ref{fig:seq_compare}), i.e. when the modulus of the temperature difference $\Delta T = T_\text{L}-T_\text{R}$ is smaller than the average lead temperature $\bar{T} = (T_\text{L} + T_\text{R})/2$, and non-linear response (Fig.~\ref{fig:nonlin_compare}).
In both cases, using the model detailed above, we could accurately reproduce the experimental data without any free parameter. The system parameters $E_C=100 \ \mathrm{\mu eV} \approx k_B \times 1.16$ K and $R_\text{L}=R_\text{R}=26 \ \mathrm{k\Omega}$ are independently extracted from charge current measurements. Figures~\ref{fig:seq_compare} and \ref{fig:nonlin_compare}a) present the same qualitative behavior, namely a periodic oscillation
of the thermovoltage with the gate-induced charge $n_g$ and a linear dependence around degeneracy points, but they exhibit different amplitudes (note that the sign of $V_{\text{th}}$ is opposite in the two cases since the temperature biases are opposite).

We first analyze the linear response regime by choosing the set of data obtained when the temperature of the leads is $T_\text{L}=134 \ \mathrm{mK}$ and $T_\text{R}=190 \ \mathrm{mK}$, such that $|\Delta T| < \bar{T}$.
In Fig.~\ref{fig:seq_compare} we compare the measured $V_{\text{th}}$ (blue dots) as a function of $n_g$ with different theoretical models. The red thin curve represents the typical sawtooth behavior which is predicted within linear response accounting only for sequential tunneling and two charge states. This is characterized by a linear function of $n_g$, crossing zero at the degeneracy points with slope $E_C\Delta T/\bar{T}$ \cite{beenakker1992}. The other two curves (red dashed and green solid) are instead determined by computing $V_{\text{th}}$ using Eq.~(\ref{eq:vth_def}) and assuming that $T_\text{I} = \bar{T}$ [see Eq.~(\ref{eq:ti_lin})]. The red dashed curve, which only accounts for sequential tunneling, shows a smoothened sawtooth behavior as a consequence of including multiple charge states in the master equation and of a finite temperature. However, both models based on sequential tunneling (thin and dashed red curves) approximately fit the experimental data only near the degeneracy points (near half integer values of $n_g$). In this case, indeed, sequential tunneling is allowed and thus dominates over co-tunneling \cite{turek2002}. On the other hand the green solid curve, computed including co-tunneling contributions, shows a strong suppression of the thermovoltage as we move away from degeneracy points. The excellent agreement between this model and the experimental measurements pinpoints the critical dependence of the thermovoltage on inelastic co-tunneling processes.

We now move to the non-linear regime. In Fig.~\ref{fig:nonlin_compare}a) we show the measured thermovoltage as a function of $n_g$ (blue dots) compared to theoretical calculations, all of which include co-tunneling contributions. The lead temperatures are $T_\text{L}=342 \ \mathrm{mK}$ and $T_\text{R}=63 \ \mathrm{mK}$, such that $|\Delta T| > \bar{T}$. 
The red dashed-dotted curve is computed within the linear response regime choosing the average lead temperature $\bar{T}$ as the characteristic temperature. More precisely, we solve Eq.~(\ref{eq:vth_def}) setting $T_\text{I} = \bar{T}$ and choosing a small temperature difference of the leads $\delta T$ around $\bar{T}$ to find the thermopower $S \equiv V_{\text{th}}/\delta T$ for $\delta T\to 0$. We then calculate the thermovoltage as $V_{\text{th}} = S(T_\text{L}-T_\text{R})$, where now $T_\text{L}=342$ mK and $T_\text{R}=63$ mK are the actual lead temperatures. As we can see from Fig.~\ref{fig:nonlin_compare}a), this linear response model overestimates the thermovoltage almost by a factor two. 
A non-linear calculation (green dashed curve) improves the agreement with the experimental data. This calculation is performed by solving Eq.~(\ref{eq:vth_def}) using the actual lead temperatures and, as before, we fix the island temperature at $T_\text{I} = \bar{T}$. 
The difference between the red dashed-dotted and green dashed curves proves that we are indeed in the non-linear response regime, and it shows that the main effect of the nonlinear response is to decrease the amplitude of the thermovoltage. 
However, we still do not obtain a good agreement with the experimental data. 

We find that we can get a perfect agreement with the experimental data if we further improve the model by determining also the island temperature $T_\text{I}$ through a heat balance equation, rather than fixing it at $\bar{T}$. 
More precisely [see Fig.~\ref{fig:setup}b)], we denote by $\dot{Q}_{\text{tun}}$ the heat current entering the island from sequential and co-tunneling events (see Supplemental Material for details \cite{suppl}) and by $Q_{\text{el-ph}} = \Sigma \mathcal{V} (T_\text{I}^5-T_\text{R}^5)$ the heat current flowing from electrons in the island to the phonons (we assume that the electronic temperature $T_\text{R}$ in the bulky right electrode is equal to the temperature of the phonons). $\mathcal{V}$ is the island volume and $\Sigma$ is the electron-phonon coupling constant which only depends on the material. 
The temperature of the island can thus be determined by the following heat balance equation
\begin{equation}
	\dot{Q}_{\text{tun}} = \dot{Q}_{\text{el-ph}}.
	\label{eq:heat_bal}
\end{equation}
The values of the parameters entering $Q_{\text{el-ph}}$ that we use are determined independently: $\mathcal{V}= 225 \times 100\times 29 \ \mathrm{nm^3}$ is estimated from SEM images and
$\Sigma$ is obtained from Ref.~\cite{dutta2017} for this device (sample B). The value,
 $\Sigma = 2.8 \, \mathrm{WK^{-5} m^{-3}}$, is close to the standard literature value for copper \cite{giazotto2006} and in agreement with measurements of other samples fabricated using the same Cu target.

The black curve in Fig.~\ref{fig:nonlin_compare}a) is thus determined by computing both $V_{\text{th}}$ and $T_\text{I}$ simultaneously by solving Eqs.~(\ref{eq:vth_def}) and (\ref{eq:heat_bal}) without any free parameters for each value of $n_g$.  As we can see, the non-linear model, complemented with the heat balance equation, is in excellent agreement with the experimental measurements, demonstrating that $T_\text{I}$ is indeed an important parameter in the non-linear regime. Conversely we have verified that, using the parameters of Fig.~\ref{fig:seq_compare} which are within the linear response regime, $V_\text{th}$ only weakly depends on the particular choice of $T_\text{I}$ between $T_\text{L}$ and $T_\text{R}$.  
In Fig.~\ref{fig:nonlin_compare}b) we plot the island temperature $T_\text{I}$, as a function of $n_g$ over a single period, determined in the same calculation that leads to the black curve in Fig.~\ref{fig:nonlin_compare}a).
Remarkably, despite the very low phonon temperature ($63 \, \mathrm{mK}$), the calculated $T_{\text{I}} \approx 250$ mK is much larger than the average lead temperature $\bar{T} = 202.5\,\mathrm{mK}$. This means that while the net charge current across the SET is zero, the heat current due to electrons tunneling back and forth is overheating the island to a temperature that is significantly larger than the average temperature, resulting in a further decrease of the thermovoltage. This is another signature of the non-linear response of the system, as it violates Eq.~(\ref{eq:ti_lin}). We further find that the island temperature displays a weak $n_g$ modulation of approximately $10 \ \mathrm{mK}$, but this prediction cannot be confirmed in the present experiment. 

\begin{figure}[!tb]
	\centering
	\includegraphics[width=1\columnwidth]{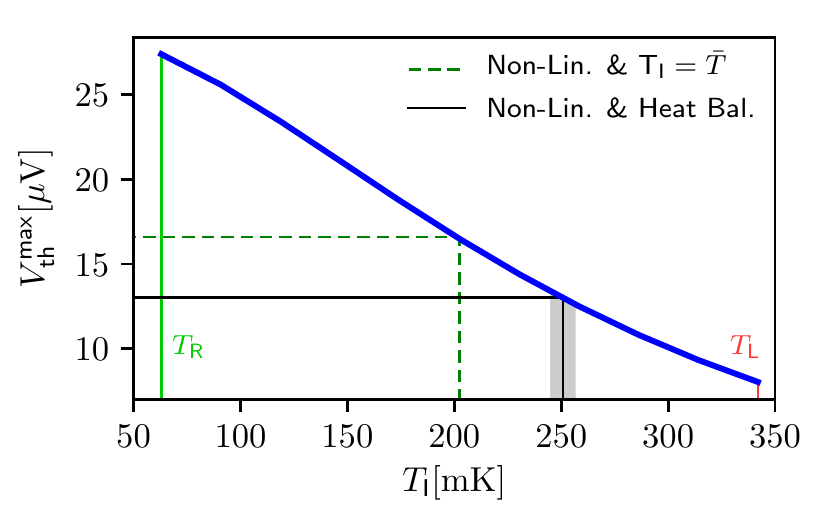}
	\caption{The maximum amplitude of the thermovoltage $V_{\text{th}}^{\text{max}}$ is plotted as a function of the island temperature, for $T_\text{R} \leq T_\text{I} \leq T_\text{L}$. The green dashed lines point to the values of $V_{\text{th}}^{\text{max}}$ and $T_\text{I}$  found in the non-linear calculation 
	at fixed $T_\text{I}=\bar{T}$ (see the green dashed curve of Fig.~\ref{fig:nonlin_compare}a) while the black solid lines and the gray area refer to the non-linear calculation including the heat balance equation (see the black solid curve of Fig.~\ref{fig:nonlin_compare}).}
	\label{fig:vmax_ti}
\end{figure}  
Finally we discuss how the thermovoltage depends on $T_\text{I}$. 
In Fig.~\ref{fig:vmax_ti} we plot $V_{\text{th}}^{\text{max}}$, the maximum amplitude of $V_{\text{th}}$, computed by solving Eq.~(\ref{eq:vth_def}) at fixed lead temperatures $T_\text{L}=342$ mK and $T_\text{R}= 63$ mK and varying $T_\text{I}$ between the lead temperatures. The black solid lines and the gray area point to the actual experimental value of $V_{\text{th}}^{\text{max}}$ and to the corresponding computed $T_\text{I}$ which differs from $\bar{T}$ [see black curves in Figs.~\ref{fig:nonlin_compare}a) and \ref{fig:nonlin_compare}b)], while the dashed green lines point to $V_{\text{th}}^{\text{max}}$  calculated setting $T_\text{I} = \bar{T}$ [see the green dashed curve in Fig.~\ref{fig:nonlin_compare}a)]. 
We find that $V_{\text{th}}^{\text{max}}$ strongly depends on the choice of $T_\text{I}$ and that it increases as $T_\text{I}$ is lowered. Indeed, at $T_\text{I} = T_\text{R} = 63$ mK, the amplitude of the thermovoltage reaches $27\, \mathrm{\mu eV}$, twice the experimental value [see blue dots in Fig.~\ref{fig:nonlin_compare}a)]. 
Thus, by increasing the energy exchange between the electrons and phonons in the island, for example by increasing the island's volume, we can lower the temperature of the island which in turn results in an increase of $V_{\text{th}}$.

{\it Conclusions.}---We performed measurements of thermovoltage in a metallic island tunnel coupled to normal leads.
Within the linear regime we nail down the role of co-tunneling in determining the thermovoltage.
Within the non-linear response regime we explore temperature biases, determined with on-chip thermometers, even larger than the average lead temperature. 
Using a theoretical model which accounts for co-tunneling and non-linear effects, we find an accurate agreement with the experimental data without any free parameters.
 In particular, we find that the temperature of the metallic island becomes an important parameter which must be determined by solving a heat balance equation for the island. Surprisingly, even if the net charge current through the system is vanishing and the coupling to the leads is symmetric, the metallic island overheats to a temperature larger than the average lead temperature. As a consequence, the amplitude of the thermovoltage oscillations decreases.

%

{\it Acknowledgments.}---This work has been supported
by the Academy of Finland (grant 312057),
by the European Union's Horizon 2020 research and innovation programme under the European Research Council (ERC) programme (grant agreement 742559),
by SNS-WIS joint lab ``QUANTRA'', 
by the SNS internal projects ``Thermoelectricity in nano-devices'', 
by the CNR-CONICET cooperation programme ``Energy conversion in quantum, nanoscale, hybrid devices'',
and by the COST ActionMP1209 ``Thermodynamics in the quantum regime''.
BD acknowledges support from the Nanosciences Fundation under the auspices of the Université Grenoble Alpes Foundation.


\begin{thebibliography}{38}%
\makeatletter
\providecommand \@ifxundefined [1]{%
 \@ifx{#1\undefined}
}%
\providecommand \@ifnum [1]{%
 \ifnum #1\expandafter \@firstoftwo
 \else \expandafter \@secondoftwo
 \fi
}%
\providecommand \@ifx [1]{%
 \ifx #1\expandafter \@firstoftwo
 \else \expandafter \@secondoftwo
 \fi
}%
\providecommand \natexlab [1]{#1}%
\providecommand \enquote  [1]{``#1''}%
\providecommand \bibnamefont  [1]{#1}%
\providecommand \bibfnamefont [1]{#1}%
\providecommand \citenamefont [1]{#1}%
\providecommand \href@noop [0]{\@secondoftwo}%
\providecommand \href [0]{\begingroup \@sanitize@url \@href}%
\providecommand \@href[1]{\@@startlink{#1}\@@href}%
\providecommand \@@href[1]{\endgroup#1\@@endlink}%
\providecommand \@sanitize@url [0]{\catcode `\\12\catcode `\$12\catcode
  `\&12\catcode `\#12\catcode `\^12\catcode `\_12\catcode `\%12\relax}%
\providecommand \@@startlink[1]{}%
\providecommand \@@endlink[0]{}%
\providecommand \url  [0]{\begingroup\@sanitize@url \@url }%
\providecommand \@url [1]{\endgroup\@href {#1}{\urlprefix }}%
\providecommand \urlprefix  [0]{URL }%
\providecommand \Eprint [0]{\href }%
\providecommand \doibase [0]{http://dx.doi.org/}%
\providecommand \selectlanguage [0]{\@gobble}%
\providecommand \bibinfo  [0]{\@secondoftwo}%
\providecommand \bibfield  [0]{\@secondoftwo}%
\providecommand \translation [1]{[#1]}%
\providecommand \BibitemOpen [0]{}%
\providecommand \bibitemStop [0]{}%
\providecommand \bibitemNoStop [0]{.\EOS\space}%
\providecommand \EOS [0]{\spacefactor3000\relax}%
\providecommand \BibitemShut  [1]{\csname bibitem#1\endcsname}%
\let\auto@bib@innerbib\@empty
\bibitem [{\citenamefont {Benenti}\ \emph {et~al.}(2017)\citenamefont
  {Benenti}, \citenamefont {Casati}, \citenamefont {Saito},\ and\ \citenamefont
  {Whitney}}]{benenti2017}%
  \BibitemOpen
  \bibfield  {author} {\bibinfo {author} {\bibfnamefont {G.}~\bibnamefont
  {Benenti}}, \bibinfo {author} {\bibfnamefont {G.}~\bibnamefont {Casati}},
  \bibinfo {author} {\bibfnamefont {K.}~\bibnamefont {Saito}}, \ and\ \bibinfo
  {author} {\bibfnamefont {R.~S.}\ \bibnamefont {Whitney}},\ }\href {\doibase
  10.1016/j.physrep.2017.05.008} {\bibfield  {journal} {\bibinfo  {journal}
  {Sci. Rep.}\ }\textbf {\bibinfo {volume} {694}},\ \bibinfo {pages} {1 }
  (\bibinfo {year} {2017})}\BibitemShut {NoStop}%
\bibitem [{\citenamefont {Molenkamp}\ \emph {et~al.}(1992)\citenamefont
  {Molenkamp}, \citenamefont {Gravier}, \citenamefont {van Houten},
  \citenamefont {Buijk}, \citenamefont {Mabesoone},\ and\ \citenamefont
  {Foxon}}]{molenkamp1992}%
  \BibitemOpen
  \bibfield  {author} {\bibinfo {author} {\bibfnamefont {L.~W.}\ \bibnamefont
  {Molenkamp}}, \bibinfo {author} {\bibfnamefont {T.}~\bibnamefont {Gravier}},
  \bibinfo {author} {\bibfnamefont {H.}~\bibnamefont {van Houten}}, \bibinfo
  {author} {\bibfnamefont {O.~J.~A.}\ \bibnamefont {Buijk}}, \bibinfo {author}
  {\bibfnamefont {M.~A.~A.}\ \bibnamefont {Mabesoone}}, \ and\ \bibinfo
  {author} {\bibfnamefont {C.~T.}\ \bibnamefont {Foxon}},\ }\href {\doibase
  10.1103/PhysRevLett.68.3765} {\bibfield  {journal} {\bibinfo  {journal}
  {Phys. Rev. Lett.}\ }\textbf {\bibinfo {volume} {68}},\ \bibinfo {pages}
  {3765} (\bibinfo {year} {1992})}\BibitemShut {NoStop}%
\bibitem [{\citenamefont {Schwab}\ \emph {et~al.}(2000)\citenamefont {Schwab},
  \citenamefont {Henriksen}, \citenamefont {Worlock},\ and\ \citenamefont
  {Roukes}}]{schwab2000}%
  \BibitemOpen
  \bibfield  {author} {\bibinfo {author} {\bibfnamefont {K.}~\bibnamefont
  {Schwab}}, \bibinfo {author} {\bibfnamefont {E.}~\bibnamefont {Henriksen}},
  \bibinfo {author} {\bibfnamefont {J.}~\bibnamefont {Worlock}}, \ and\
  \bibinfo {author} {\bibfnamefont {M.~L.}\ \bibnamefont {Roukes}},\ }\href
  {\doibase 10.1038/35010065} {\bibfield  {journal} {\bibinfo  {journal}
  {Nature}\ }\textbf {\bibinfo {volume} {404}},\ \bibinfo {pages} {974}
  (\bibinfo {year} {2000})}\BibitemShut {NoStop}%
\bibitem [{\citenamefont {Chiatti}\ \emph {et~al.}(2006)\citenamefont
  {Chiatti}, \citenamefont {Nicholls}, \citenamefont {Proskuryakov},
  \citenamefont {Lumpkin}, \citenamefont {Farrer},\ and\ \citenamefont
  {Ritchie}}]{chiatti2006}%
  \BibitemOpen
  \bibfield  {author} {\bibinfo {author} {\bibfnamefont {O.}~\bibnamefont
  {Chiatti}}, \bibinfo {author} {\bibfnamefont {J.~T.}\ \bibnamefont
  {Nicholls}}, \bibinfo {author} {\bibfnamefont {Y.~Y.}\ \bibnamefont
  {Proskuryakov}}, \bibinfo {author} {\bibfnamefont {N.}~\bibnamefont
  {Lumpkin}}, \bibinfo {author} {\bibfnamefont {I.}~\bibnamefont {Farrer}}, \
  and\ \bibinfo {author} {\bibfnamefont {D.~A.}\ \bibnamefont {Ritchie}},\
  }\href {\doibase 10.1103/PhysRevLett.97.056601} {\bibfield  {journal}
  {\bibinfo  {journal} {Phys. Rev. Lett.}\ }\textbf {\bibinfo {volume} {97}},\
  \bibinfo {pages} {056601} (\bibinfo {year} {2006})}\BibitemShut {NoStop}%
\bibitem [{\citenamefont {Meschke}\ \emph {et~al.}(2006)\citenamefont
  {Meschke}, \citenamefont {Guichard},\ and\ \citenamefont
  {Pekola}}]{meschke2006}%
  \BibitemOpen
  \bibfield  {author} {\bibinfo {author} {\bibfnamefont {M.}~\bibnamefont
  {Meschke}}, \bibinfo {author} {\bibfnamefont {W.}~\bibnamefont {Guichard}}, \
  and\ \bibinfo {author} {\bibfnamefont {J.~P.}\ \bibnamefont {Pekola}},\
  }\href {\doibase 10.1038/nature05276} {\bibfield  {journal} {\bibinfo
  {journal} {Nature}\ }\textbf {\bibinfo {volume} {444}},\ \bibinfo {pages}
  {187} (\bibinfo {year} {2006})}\BibitemShut {NoStop}%
\bibitem [{\citenamefont {Hoffmann}\ \emph {et~al.}(2009)\citenamefont
  {Hoffmann}, \citenamefont {Nilsson}, \citenamefont {Matthews}, \citenamefont
  {Nakpathomkun}, \citenamefont {Persson}, \citenamefont {Samuelson},\ and\
  \citenamefont {Linke}}]{hoffmann2009}%
  \BibitemOpen
  \bibfield  {author} {\bibinfo {author} {\bibfnamefont {E.~A.}\ \bibnamefont
  {Hoffmann}}, \bibinfo {author} {\bibfnamefont {H.~A.}\ \bibnamefont
  {Nilsson}}, \bibinfo {author} {\bibfnamefont {J.~E.}\ \bibnamefont
  {Matthews}}, \bibinfo {author} {\bibfnamefont {N.}~\bibnamefont
  {Nakpathomkun}}, \bibinfo {author} {\bibfnamefont {A.~I.}\ \bibnamefont
  {Persson}}, \bibinfo {author} {\bibfnamefont {L.}~\bibnamefont {Samuelson}},
  \ and\ \bibinfo {author} {\bibfnamefont {H.}~\bibnamefont {Linke}},\ }\href
  {\doibase 10.1021/nl8034042} {\bibfield  {journal} {\bibinfo  {journal} {Nano
  Lett.}\ }\textbf {\bibinfo {volume} {9}},\ \bibinfo {pages} {779} (\bibinfo
  {year} {2009})}\BibitemShut {NoStop}%
\bibitem [{\citenamefont {Jezouin}\ \emph {et~al.}(2013)\citenamefont
  {Jezouin}, \citenamefont {Parmentier}, \citenamefont {Anthore}, \citenamefont
  {Gennser}, \citenamefont {Cavanna}, \citenamefont {Jin},\ and\ \citenamefont
  {Pierre}}]{jezouin2013}%
  \BibitemOpen
  \bibfield  {author} {\bibinfo {author} {\bibfnamefont {S.}~\bibnamefont
  {Jezouin}}, \bibinfo {author} {\bibfnamefont {F.~D.}\ \bibnamefont
  {Parmentier}}, \bibinfo {author} {\bibfnamefont {A.}~\bibnamefont {Anthore}},
  \bibinfo {author} {\bibfnamefont {U.}~\bibnamefont {Gennser}}, \bibinfo
  {author} {\bibfnamefont {A.}~\bibnamefont {Cavanna}}, \bibinfo {author}
  {\bibfnamefont {Y.}~\bibnamefont {Jin}}, \ and\ \bibinfo {author}
  {\bibfnamefont {F.}~\bibnamefont {Pierre}},\ }\href {\doibase
  10.1126/science.1241912} {\bibfield  {journal} {\bibinfo  {journal}
  {Science}\ }\textbf {\bibinfo {volume} {342}},\ \bibinfo {pages} {601}
  (\bibinfo {year} {2013})}\BibitemShut {NoStop}%
\bibitem [{\citenamefont {Banerjee}\ \emph {et~al.}(2017)\citenamefont
  {Banerjee}, \citenamefont {Heiblum}, \citenamefont {Rosenblatt},
  \citenamefont {Oreg}, \citenamefont {Feldman}, \citenamefont {Stern},\ and\
  \citenamefont {Umansky}}]{banerjee2017}%
  \BibitemOpen
  \bibfield  {author} {\bibinfo {author} {\bibfnamefont {M.}~\bibnamefont
  {Banerjee}}, \bibinfo {author} {\bibfnamefont {M.}~\bibnamefont {Heiblum}},
  \bibinfo {author} {\bibfnamefont {A.}~\bibnamefont {Rosenblatt}}, \bibinfo
  {author} {\bibfnamefont {Y.}~\bibnamefont {Oreg}}, \bibinfo {author}
  {\bibfnamefont {D.~E.}\ \bibnamefont {Feldman}}, \bibinfo {author}
  {\bibfnamefont {A.}~\bibnamefont {Stern}}, \ and\ \bibinfo {author}
  {\bibfnamefont {V.}~\bibnamefont {Umansky}},\ }\href {\doibase
  10.1038/nature22052} {\bibfield  {journal} {\bibinfo  {journal} {Nature}\
  }\textbf {\bibinfo {volume} {545}},\ \bibinfo {pages} {75} (\bibinfo {year}
  {2017})}\BibitemShut {NoStop}%
\bibitem [{\citenamefont {Dutta}\ \emph {et~al.}(2017)\citenamefont {Dutta},
  \citenamefont {Peltonen}, \citenamefont {Antonenko}, \citenamefont {Meschke},
  \citenamefont {Skvortsov}, \citenamefont {Kubala}, \citenamefont {K\"onig},
  \citenamefont {Winkelmann}, \citenamefont {Courtois},\ and\ \citenamefont
  {Pekola}}]{dutta2017}%
  \BibitemOpen
  \bibfield  {author} {\bibinfo {author} {\bibfnamefont {B.}~\bibnamefont
  {Dutta}}, \bibinfo {author} {\bibfnamefont {J.~T.}\ \bibnamefont {Peltonen}},
  \bibinfo {author} {\bibfnamefont {D.~S.}\ \bibnamefont {Antonenko}}, \bibinfo
  {author} {\bibfnamefont {M.}~\bibnamefont {Meschke}}, \bibinfo {author}
  {\bibfnamefont {M.~A.}\ \bibnamefont {Skvortsov}}, \bibinfo {author}
  {\bibfnamefont {B.}~\bibnamefont {Kubala}}, \bibinfo {author} {\bibfnamefont
  {J.}~\bibnamefont {K\"onig}}, \bibinfo {author} {\bibfnamefont {C.~B.}\
  \bibnamefont {Winkelmann}}, \bibinfo {author} {\bibfnamefont
  {H.}~\bibnamefont {Courtois}}, \ and\ \bibinfo {author} {\bibfnamefont
  {J.~P.}\ \bibnamefont {Pekola}},\ }\href {\doibase
  10.1103/PhysRevLett.119.077701} {\bibfield  {journal} {\bibinfo  {journal}
  {Phys. Rev. Lett.}\ }\textbf {\bibinfo {volume} {119}},\ \bibinfo {pages}
  {077701} (\bibinfo {year} {2017})}\BibitemShut {NoStop}%
\bibitem [{\citenamefont {Cui}\ \emph {et~al.}(2017)\citenamefont {Cui},
  \citenamefont {Jeong}, \citenamefont {Hur}, \citenamefont {Matt},
  \citenamefont {Kl{\"o}ckner}, \citenamefont {Pauly}, \citenamefont {Nielaba},
  \citenamefont {Cuevas}, \citenamefont {Meyhofer},\ and\ \citenamefont
  {Reddy}}]{cui2017}%
  \BibitemOpen
  \bibfield  {author} {\bibinfo {author} {\bibfnamefont {L.}~\bibnamefont
  {Cui}}, \bibinfo {author} {\bibfnamefont {W.}~\bibnamefont {Jeong}}, \bibinfo
  {author} {\bibfnamefont {S.}~\bibnamefont {Hur}}, \bibinfo {author}
  {\bibfnamefont {M.}~\bibnamefont {Matt}}, \bibinfo {author} {\bibfnamefont
  {J.~C.}\ \bibnamefont {Kl{\"o}ckner}}, \bibinfo {author} {\bibfnamefont
  {F.}~\bibnamefont {Pauly}}, \bibinfo {author} {\bibfnamefont
  {P.}~\bibnamefont {Nielaba}}, \bibinfo {author} {\bibfnamefont {J.~C.}\
  \bibnamefont {Cuevas}}, \bibinfo {author} {\bibfnamefont {E.}~\bibnamefont
  {Meyhofer}}, \ and\ \bibinfo {author} {\bibfnamefont {P.}~\bibnamefont
  {Reddy}},\ }\href {\doibase 10.1126/science.aam6622} {\bibfield  {journal}
  {\bibinfo  {journal} {Science}\ }\textbf {\bibinfo {volume} {355}},\ \bibinfo
  {pages} {1192} (\bibinfo {year} {2017})}\BibitemShut {NoStop}%
\bibitem [{\citenamefont {Mosso}\ \emph {et~al.}(2017)\citenamefont {Mosso},
  \citenamefont {Drechsler}, \citenamefont {Menges}, \citenamefont {Nirmalraj},
  \citenamefont {Karg}, \citenamefont {Riel},\ and\ \citenamefont
  {Gotsmann}}]{mosso2017}%
  \BibitemOpen
  \bibfield  {author} {\bibinfo {author} {\bibfnamefont {N.}~\bibnamefont
  {Mosso}}, \bibinfo {author} {\bibfnamefont {U.}~\bibnamefont {Drechsler}},
  \bibinfo {author} {\bibfnamefont {F.}~\bibnamefont {Menges}}, \bibinfo
  {author} {\bibfnamefont {P.}~\bibnamefont {Nirmalraj}}, \bibinfo {author}
  {\bibfnamefont {S.}~\bibnamefont {Karg}}, \bibinfo {author} {\bibfnamefont
  {H.}~\bibnamefont {Riel}}, \ and\ \bibinfo {author} {\bibfnamefont
  {B.}~\bibnamefont {Gotsmann}},\ }\href {\doibase 10.1038/nnano.2016.302}
  {\bibfield  {journal} {\bibinfo  {journal} {Nat. Nanotechnol.}\ }\textbf
  {\bibinfo {volume} {12}},\ \bibinfo {pages} {430} (\bibinfo {year}
  {2017})}\BibitemShut {NoStop}%
\bibitem [{\citenamefont {Roddaro}\ \emph {et~al.}(2013)\citenamefont
  {Roddaro}, \citenamefont {Ercolani}, \citenamefont {Safeen}, \citenamefont
  {Suomalainen}, \citenamefont {Rossella}, \citenamefont {Giazotto},
  \citenamefont {Sorba},\ and\ \citenamefont {Beltram}}]{roddaro2013}%
  \BibitemOpen
  \bibfield  {author} {\bibinfo {author} {\bibfnamefont {S.}~\bibnamefont
  {Roddaro}}, \bibinfo {author} {\bibfnamefont {D.}~\bibnamefont {Ercolani}},
  \bibinfo {author} {\bibfnamefont {M.~A.}\ \bibnamefont {Safeen}}, \bibinfo
  {author} {\bibfnamefont {S.}~\bibnamefont {Suomalainen}}, \bibinfo {author}
  {\bibfnamefont {F.}~\bibnamefont {Rossella}}, \bibinfo {author}
  {\bibfnamefont {F.}~\bibnamefont {Giazotto}}, \bibinfo {author}
  {\bibfnamefont {L.}~\bibnamefont {Sorba}}, \ and\ \bibinfo {author}
  {\bibfnamefont {F.}~\bibnamefont {Beltram}},\ }\href {\doibase
  10.1021/nl401482p} {\bibfield  {journal} {\bibinfo  {journal} {Nano Lett.}\
  }\textbf {\bibinfo {volume} {13}},\ \bibinfo {pages} {3638} (\bibinfo {year}
  {2013})}\BibitemShut {NoStop}%
\bibitem [{\citenamefont {Wu}\ \emph {et~al.}(2013)\citenamefont {Wu},
  \citenamefont {Gooth}, \citenamefont {Zianni}, \citenamefont {Svensson},
  \citenamefont {Gluschke}, \citenamefont {Dick}, \citenamefont {Thelander},
  \citenamefont {Nielsch},\ and\ \citenamefont {Linke}}]{wu2013}%
  \BibitemOpen
  \bibfield  {author} {\bibinfo {author} {\bibfnamefont {P.~M.}\ \bibnamefont
  {Wu}}, \bibinfo {author} {\bibfnamefont {J.}~\bibnamefont {Gooth}}, \bibinfo
  {author} {\bibfnamefont {X.}~\bibnamefont {Zianni}}, \bibinfo {author}
  {\bibfnamefont {S.~F.}\ \bibnamefont {Svensson}}, \bibinfo {author}
  {\bibfnamefont {J.~G.}\ \bibnamefont {Gluschke}}, \bibinfo {author}
  {\bibfnamefont {K.~A.}\ \bibnamefont {Dick}}, \bibinfo {author}
  {\bibfnamefont {C.}~\bibnamefont {Thelander}}, \bibinfo {author}
  {\bibfnamefont {K.}~\bibnamefont {Nielsch}}, \ and\ \bibinfo {author}
  {\bibfnamefont {H.}~\bibnamefont {Linke}},\ }\href {\doibase
  10.1021/nl401501j} {\bibfield  {journal} {\bibinfo  {journal} {Nano Lett.}\
  }\textbf {\bibinfo {volume} {13}},\ \bibinfo {pages} {4080} (\bibinfo {year}
  {2013})}\BibitemShut {NoStop}%
\bibitem [{\citenamefont {Dzurak}\ \emph {et~al.}(1993)\citenamefont {Dzurak},
  \citenamefont {Smith}, \citenamefont {Pepper}, \citenamefont {Ritchie},
  \citenamefont {Frost}, \citenamefont {Jones},\ and\ \citenamefont
  {Hasko}}]{dzurak1993}%
  \BibitemOpen
  \bibfield  {author} {\bibinfo {author} {\bibfnamefont {A.~S.}\ \bibnamefont
  {Dzurak}}, \bibinfo {author} {\bibfnamefont {C.~G.}\ \bibnamefont {Smith}},
  \bibinfo {author} {\bibfnamefont {M.}~\bibnamefont {Pepper}}, \bibinfo
  {author} {\bibfnamefont {D.}~\bibnamefont {Ritchie}}, \bibinfo {author}
  {\bibfnamefont {J.}~\bibnamefont {Frost}}, \bibinfo {author} {\bibfnamefont
  {G.}~\bibnamefont {Jones}}, \ and\ \bibinfo {author} {\bibfnamefont
  {D.}~\bibnamefont {Hasko}},\ }\href {\doibase 10.1016/0038-1098(93)90819-9}
  {\bibfield  {journal} {\bibinfo  {journal} {Solid State Commun.}\ }\textbf
  {\bibinfo {volume} {87}},\ \bibinfo {pages} {1145} (\bibinfo {year}
  {1993})}\BibitemShut {NoStop}%
\bibitem [{\citenamefont {Staring}\ \emph {et~al.}(1993)\citenamefont
  {Staring}, \citenamefont {Molenkamp}, \citenamefont {Alphenaar},
  \citenamefont {van Houten}, \citenamefont {Buyk}, \citenamefont {Mabesoone},
  \citenamefont {Beenakker},\ and\ \citenamefont {Foxon}}]{staring1993}%
  \BibitemOpen
  \bibfield  {author} {\bibinfo {author} {\bibfnamefont {A.~A.~M.}\
  \bibnamefont {Staring}}, \bibinfo {author} {\bibfnamefont {L.~W.}\
  \bibnamefont {Molenkamp}}, \bibinfo {author} {\bibfnamefont {B.~W.}\
  \bibnamefont {Alphenaar}}, \bibinfo {author} {\bibfnamefont {H.}~\bibnamefont
  {van Houten}}, \bibinfo {author} {\bibfnamefont {O.~J.~A.}\ \bibnamefont
  {Buyk}}, \bibinfo {author} {\bibfnamefont {M.~A.~A.}\ \bibnamefont
  {Mabesoone}}, \bibinfo {author} {\bibfnamefont {C.~W.~J.}\ \bibnamefont
  {Beenakker}}, \ and\ \bibinfo {author} {\bibfnamefont {C.~T.}\ \bibnamefont
  {Foxon}},\ }\href {\doibase 10.1209/0295-5075/22/1/011} {\bibfield  {journal}
  {\bibinfo  {journal} {EPL}\ }\textbf {\bibinfo {volume} {22}},\ \bibinfo
  {pages} {57} (\bibinfo {year} {1993})}\BibitemShut {NoStop}%
\bibitem [{\citenamefont {Molenkamp}\ \emph {et~al.}(1994)\citenamefont
  {Molenkamp}, \citenamefont {Staring}, \citenamefont {Alphenaar},
  \citenamefont {van Houten},\ and\ \citenamefont {Beenakker}}]{molenkamp1994}%
  \BibitemOpen
  \bibfield  {author} {\bibinfo {author} {\bibfnamefont {L.}~\bibnamefont
  {Molenkamp}}, \bibinfo {author} {\bibfnamefont {A.~A.~M.}\ \bibnamefont
  {Staring}}, \bibinfo {author} {\bibfnamefont {B.~W.}\ \bibnamefont
  {Alphenaar}}, \bibinfo {author} {\bibfnamefont {H.}~\bibnamefont {van
  Houten}}, \ and\ \bibinfo {author} {\bibfnamefont {C.~W.~J.}\ \bibnamefont
  {Beenakker}},\ }\href {\doibase 10.1088/0268-1242/9/5S/136} {\bibfield
  {journal} {\bibinfo  {journal} {Semicond. Sci. Technol.}\ }\textbf {\bibinfo
  {volume} {9}},\ \bibinfo {pages} {903} (\bibinfo {year} {1994})}\BibitemShut
  {NoStop}%
\bibitem [{\citenamefont {Dzurak}\ \emph {et~al.}(1997)\citenamefont {Dzurak},
  \citenamefont {Smith}, \citenamefont {Barnes}, \citenamefont {Pepper},
  \citenamefont {Mart\'{\i}n-Moreno}, \citenamefont {Liang}, \citenamefont
  {Ritchie},\ and\ \citenamefont {Jones}}]{dzurak1997}%
  \BibitemOpen
  \bibfield  {author} {\bibinfo {author} {\bibfnamefont {A.~S.}\ \bibnamefont
  {Dzurak}}, \bibinfo {author} {\bibfnamefont {C.~G.}\ \bibnamefont {Smith}},
  \bibinfo {author} {\bibfnamefont {C.~H.~W.}\ \bibnamefont {Barnes}}, \bibinfo
  {author} {\bibfnamefont {M.}~\bibnamefont {Pepper}}, \bibinfo {author}
  {\bibfnamefont {L.}~\bibnamefont {Mart\'{\i}n-Moreno}}, \bibinfo {author}
  {\bibfnamefont {C.~T.}\ \bibnamefont {Liang}}, \bibinfo {author}
  {\bibfnamefont {D.~A.}\ \bibnamefont {Ritchie}}, \ and\ \bibinfo {author}
  {\bibfnamefont {G.~A.~C.}\ \bibnamefont {Jones}},\ }\href {\doibase
  10.1103/PhysRevB.55.R10197} {\bibfield  {journal} {\bibinfo  {journal} {Phys.
  Rev. B}\ }\textbf {\bibinfo {volume} {55}},\ \bibinfo {pages} {R10197}
  (\bibinfo {year} {1997})}\BibitemShut {NoStop}%
\bibitem [{\citenamefont {M\"oller}\ \emph {et~al.}(1998)\citenamefont
  {M\"oller}, \citenamefont {Buhmann}, \citenamefont {Godijn},\ and\
  \citenamefont {Molenkamp}}]{moller1998}%
  \BibitemOpen
  \bibfield  {author} {\bibinfo {author} {\bibfnamefont {S.}~\bibnamefont
  {M\"oller}}, \bibinfo {author} {\bibfnamefont {H.}~\bibnamefont {Buhmann}},
  \bibinfo {author} {\bibfnamefont {S.~F.}\ \bibnamefont {Godijn}}, \ and\
  \bibinfo {author} {\bibfnamefont {L.~W.}\ \bibnamefont {Molenkamp}},\ }\href
  {\doibase 10.1103/PhysRevLett.81.5197} {\bibfield  {journal} {\bibinfo
  {journal} {Phys. Rev. Lett.}\ }\textbf {\bibinfo {volume} {81}},\ \bibinfo
  {pages} {5197} (\bibinfo {year} {1998})}\BibitemShut {NoStop}%
\bibitem [{\citenamefont {Godijn}\ \emph {et~al.}(1999)\citenamefont {Godijn},
  \citenamefont {M\"oller}, \citenamefont {Buhmann}, \citenamefont
  {Molenkamp},\ and\ \citenamefont {van Langen}}]{godijn1999}%
  \BibitemOpen
  \bibfield  {author} {\bibinfo {author} {\bibfnamefont {S.~F.}\ \bibnamefont
  {Godijn}}, \bibinfo {author} {\bibfnamefont {S.}~\bibnamefont {M\"oller}},
  \bibinfo {author} {\bibfnamefont {H.}~\bibnamefont {Buhmann}}, \bibinfo
  {author} {\bibfnamefont {L.~W.}\ \bibnamefont {Molenkamp}}, \ and\ \bibinfo
  {author} {\bibfnamefont {S.~A.}\ \bibnamefont {van Langen}},\ }\href
  {\doibase 10.1103/PhysRevLett.82.2927} {\bibfield  {journal} {\bibinfo
  {journal} {Phys. Rev. Lett.}\ }\textbf {\bibinfo {volume} {82}},\ \bibinfo
  {pages} {2927} (\bibinfo {year} {1999})}\BibitemShut {NoStop}%
\bibitem [{\citenamefont {Llaguno}\ \emph {et~al.}(2004)\citenamefont
  {Llaguno}, \citenamefont {Fischer}, \citenamefont {Johnson},\ and\
  \citenamefont {Hone}}]{llaguno2004}%
  \BibitemOpen
  \bibfield  {author} {\bibinfo {author} {\bibfnamefont {M.~C.}\ \bibnamefont
  {Llaguno}}, \bibinfo {author} {\bibfnamefont {J.~E.}\ \bibnamefont
  {Fischer}}, \bibinfo {author} {\bibfnamefont {A.~T.}\ \bibnamefont
  {Johnson}}, \ and\ \bibinfo {author} {\bibfnamefont {J.}~\bibnamefont
  {Hone}},\ }\href {\doibase 10.1021/nl0348488} {\bibfield  {journal} {\bibinfo
   {journal} {Nano Lett.}\ }\textbf {\bibinfo {volume} {4}},\ \bibinfo {pages}
  {45} (\bibinfo {year} {2004})}\BibitemShut {NoStop}%
\bibitem [{\citenamefont {Scheibner}\ \emph {et~al.}(2005)\citenamefont
  {Scheibner}, \citenamefont {Buhmann}, \citenamefont {Reuter}, \citenamefont
  {Kiselev},\ and\ \citenamefont {Molenkamp}}]{scheibner2005}%
  \BibitemOpen
  \bibfield  {author} {\bibinfo {author} {\bibfnamefont {R.}~\bibnamefont
  {Scheibner}}, \bibinfo {author} {\bibfnamefont {H.}~\bibnamefont {Buhmann}},
  \bibinfo {author} {\bibfnamefont {D.}~\bibnamefont {Reuter}}, \bibinfo
  {author} {\bibfnamefont {M.~N.}\ \bibnamefont {Kiselev}}, \ and\ \bibinfo
  {author} {\bibfnamefont {L.~W.}\ \bibnamefont {Molenkamp}},\ }\href {\doibase
  10.1103/PhysRevLett.95.176602} {\bibfield  {journal} {\bibinfo  {journal}
  {Phys. Rev. Lett.}\ }\textbf {\bibinfo {volume} {95}},\ \bibinfo {pages}
  {176602} (\bibinfo {year} {2005})}\BibitemShut {NoStop}%
\bibitem [{\citenamefont {Pogosov}\ \emph {et~al.}(2006)\citenamefont
  {Pogosov}, \citenamefont {Budantsev}, \citenamefont {Lavrov}, \citenamefont
  {Plotnikov}, \citenamefont {Bakarov}, \citenamefont {Toropov},\ and\
  \citenamefont {Portal}}]{pogosov2006}%
  \BibitemOpen
  \bibfield  {author} {\bibinfo {author} {\bibfnamefont {A.~G.}\ \bibnamefont
  {Pogosov}}, \bibinfo {author} {\bibfnamefont {M.~V.}\ \bibnamefont
  {Budantsev}}, \bibinfo {author} {\bibfnamefont {R.~A.}\ \bibnamefont
  {Lavrov}}, \bibinfo {author} {\bibfnamefont {A.~E.}\ \bibnamefont
  {Plotnikov}}, \bibinfo {author} {\bibfnamefont {A.~K.}\ \bibnamefont
  {Bakarov}}, \bibinfo {author} {\bibfnamefont {A.~I.}\ \bibnamefont
  {Toropov}}, \ and\ \bibinfo {author} {\bibfnamefont {J.~C.}\ \bibnamefont
  {Portal}},\ }\href {\doibase 10.1134/S002136400603009X} {\bibfield  {journal}
  {\bibinfo  {journal} {JETP Lett.}\ }\textbf {\bibinfo {volume} {83}},\
  \bibinfo {pages} {122} (\bibinfo {year} {2006})}\BibitemShut {NoStop}%
\bibitem [{\citenamefont {Scheibner}\ \emph {et~al.}(2007)\citenamefont
  {Scheibner}, \citenamefont {Novik}, \citenamefont {Borzenko}, \citenamefont
  {K\"onig}, \citenamefont {Reuter}, \citenamefont {Wieck}, \citenamefont
  {Buhmann},\ and\ \citenamefont {Molenkamp}}]{scheibner2007}%
  \BibitemOpen
  \bibfield  {author} {\bibinfo {author} {\bibfnamefont {R.}~\bibnamefont
  {Scheibner}}, \bibinfo {author} {\bibfnamefont {E.~G.}\ \bibnamefont
  {Novik}}, \bibinfo {author} {\bibfnamefont {T.}~\bibnamefont {Borzenko}},
  \bibinfo {author} {\bibfnamefont {M.}~\bibnamefont {K\"onig}}, \bibinfo
  {author} {\bibfnamefont {D.}~\bibnamefont {Reuter}}, \bibinfo {author}
  {\bibfnamefont {A.~D.}\ \bibnamefont {Wieck}}, \bibinfo {author}
  {\bibfnamefont {H.}~\bibnamefont {Buhmann}}, \ and\ \bibinfo {author}
  {\bibfnamefont {L.~W.}\ \bibnamefont {Molenkamp}},\ }\href {\doibase
  10.1103/PhysRevB.75.041301} {\bibfield  {journal} {\bibinfo  {journal} {Phys.
  Rev. B}\ }\textbf {\bibinfo {volume} {75}},\ \bibinfo {pages} {041301}
  (\bibinfo {year} {2007})}\BibitemShut {NoStop}%
\bibitem [{\citenamefont {Svensson}\ \emph {et~al.}(2012)\citenamefont
  {Svensson}, \citenamefont {Persson}, \citenamefont {Hoffmann}, \citenamefont
  {Nakpathomkun}, \citenamefont {Nilsson}, \citenamefont {Xu}, \citenamefont
  {Samuelson},\ and\ \citenamefont {Linke}}]{svensson2012}%
  \BibitemOpen
  \bibfield  {author} {\bibinfo {author} {\bibfnamefont {S.~F.}\ \bibnamefont
  {Svensson}}, \bibinfo {author} {\bibfnamefont {A.~I.}\ \bibnamefont
  {Persson}}, \bibinfo {author} {\bibfnamefont {E.~A.}\ \bibnamefont
  {Hoffmann}}, \bibinfo {author} {\bibfnamefont {N.}~\bibnamefont
  {Nakpathomkun}}, \bibinfo {author} {\bibfnamefont {H.~A.}\ \bibnamefont
  {Nilsson}}, \bibinfo {author} {\bibfnamefont {H.~Q.}\ \bibnamefont {Xu}},
  \bibinfo {author} {\bibfnamefont {L.}~\bibnamefont {Samuelson}}, \ and\
  \bibinfo {author} {\bibfnamefont {H.}~\bibnamefont {Linke}},\ }\href
  {\doibase 10.1088/1367-2630/14/3/033041} {\bibfield  {journal} {\bibinfo
  {journal} {New J. Phys.}\ }\textbf {\bibinfo {volume} {14}},\ \bibinfo
  {pages} {033041} (\bibinfo {year} {2012})}\BibitemShut {NoStop}%
\bibitem [{\citenamefont {Svensson}\ \emph {et~al.}(2013)\citenamefont
  {Svensson}, \citenamefont {Hoffmann}, \citenamefont {Nakpathomkun},
  \citenamefont {Wu}, \citenamefont {Xu}, \citenamefont {Nilsson},
  \citenamefont {S{\'a}nchez}, \citenamefont {Kashcheyevs},\ and\ \citenamefont
  {Linke}}]{svensson2013}%
  \BibitemOpen
  \bibfield  {author} {\bibinfo {author} {\bibfnamefont {S.~F.}\ \bibnamefont
  {Svensson}}, \bibinfo {author} {\bibfnamefont {E.~A.}\ \bibnamefont
  {Hoffmann}}, \bibinfo {author} {\bibfnamefont {N.}~\bibnamefont
  {Nakpathomkun}}, \bibinfo {author} {\bibfnamefont {P.~M.}\ \bibnamefont
  {Wu}}, \bibinfo {author} {\bibfnamefont {H.~Q.}\ \bibnamefont {Xu}}, \bibinfo
  {author} {\bibfnamefont {H.~A.}\ \bibnamefont {Nilsson}}, \bibinfo {author}
  {\bibfnamefont {D.}~\bibnamefont {S{\'a}nchez}}, \bibinfo {author}
  {\bibfnamefont {V.}~\bibnamefont {Kashcheyevs}}, \ and\ \bibinfo {author}
  {\bibfnamefont {H.}~\bibnamefont {Linke}},\ }\href {\doibase
  10.1088/1367-2630/15/10/105011} {\bibfield  {journal} {\bibinfo  {journal}
  {New J. Phys.}\ }\textbf {\bibinfo {volume} {15}},\ \bibinfo {pages} {105011}
  (\bibinfo {year} {2013})}\BibitemShut {NoStop}%
\bibitem{dutta2018}
B. Dutta, D. Majidi, A. G. Corral, P. Erdman, S. Florens, T. Costi, H. Courtois, and C. B. Winkelmann,
\href{https://arxiv.org/abs/1811.04219}{arXiv:1811.04219 (2018)}.
\bibitem [{\citenamefont {Beenakker}\ and\ \citenamefont
  {Staring}(1992)}]{beenakker1992}%
  \BibitemOpen
  \bibfield  {author} {\bibinfo {author} {\bibfnamefont {C.~W.~J.}\
  \bibnamefont {Beenakker}}\ and\ \bibinfo {author} {\bibfnamefont {A.~A.~M.}\
  \bibnamefont {Staring}},\ }\href {\doibase 10.1103/PhysRevB.46.9667}
  {\bibfield  {journal} {\bibinfo  {journal} {Phys. Rev. B}\ }\textbf {\bibinfo
  {volume} {46}},\ \bibinfo {pages} {9667} (\bibinfo {year}
  {1992})}\BibitemShut {NoStop}%
\bibitem [{\citenamefont {Andreev}\ and\ \citenamefont
  {Matveev}(2001)}]{andreev2001}%
  \BibitemOpen
  \bibfield  {author} {\bibinfo {author} {\bibfnamefont {A.~V.}\ \bibnamefont
  {Andreev}}\ and\ \bibinfo {author} {\bibfnamefont {K.~A.}\ \bibnamefont
  {Matveev}},\ }\href {\doibase 10.1103/PhysRevLett.86.280} {\bibfield
  {journal} {\bibinfo  {journal} {Phys. Rev. Lett.}\ }\textbf {\bibinfo
  {volume} {86}},\ \bibinfo {pages} {280} (\bibinfo {year} {2001})}\BibitemShut
  {NoStop}%
\bibitem [{\citenamefont {Turek}\ and\ \citenamefont
  {Matveev}(2002)}]{turek2002}%
  \BibitemOpen
  \bibfield  {author} {\bibinfo {author} {\bibfnamefont {M.}~\bibnamefont
  {Turek}}\ and\ \bibinfo {author} {\bibfnamefont {K.~A.}\ \bibnamefont
  {Matveev}},\ }\href {\doibase 10.1103/PhysRevB.65.115332} {\bibfield
  {journal} {\bibinfo  {journal} {Phys. Rev. B}\ }\textbf {\bibinfo {volume}
  {65}},\ \bibinfo {pages} {115332} (\bibinfo {year} {2002})}\BibitemShut
  {NoStop}%
\bibitem [{\citenamefont {Koch}\ \emph {et~al.}(2004)\citenamefont {Koch},
  \citenamefont {von Oppen}, \citenamefont {Oreg},\ and\ \citenamefont
  {Sela}}]{koch2004}%
  \BibitemOpen
  \bibfield  {author} {\bibinfo {author} {\bibfnamefont {J.}~\bibnamefont
  {Koch}}, \bibinfo {author} {\bibfnamefont {F.}~\bibnamefont {von Oppen}},
  \bibinfo {author} {\bibfnamefont {Y.}~\bibnamefont {Oreg}}, \ and\ \bibinfo
  {author} {\bibfnamefont {E.}~\bibnamefont {Sela}},\ }\href {\doibase
  10.1103/PhysRevB.70.195107} {\bibfield  {journal} {\bibinfo  {journal} {Phys.
  Rev. B}\ }\textbf {\bibinfo {volume} {70}},\ \bibinfo {pages} {195107}
  (\bibinfo {year} {2004})}\BibitemShut {NoStop}%
\bibitem [{\citenamefont {Kubala}\ and\ \citenamefont
  {K\"onig}(2006)}]{kubala2006}%
  \BibitemOpen
  \bibfield  {author} {\bibinfo {author} {\bibfnamefont {B.}~\bibnamefont
  {Kubala}}\ and\ \bibinfo {author} {\bibfnamefont {J.}~\bibnamefont
  {K\"onig}},\ }\href {\doibase 10.1103/PhysRevB.73.195316} {\bibfield
  {journal} {\bibinfo  {journal} {Phys. Rev. B}\ }\textbf {\bibinfo {volume}
  {73}},\ \bibinfo {pages} {195316} (\bibinfo {year} {2006})}\BibitemShut
  {NoStop}%
\bibitem [{\citenamefont {Kubala}\ \emph {et~al.}(2008)\citenamefont {Kubala},
  \citenamefont {K\"onig},\ and\ \citenamefont {Pekola}}]{kubala2008}%
  \BibitemOpen
  \bibfield  {author} {\bibinfo {author} {\bibfnamefont {B.}~\bibnamefont
  {Kubala}}, \bibinfo {author} {\bibfnamefont {J.}~\bibnamefont {K\"onig}}, \
  and\ \bibinfo {author} {\bibfnamefont {J.}~\bibnamefont {Pekola}},\ }\href
  {\doibase 10.1103/PhysRevLett.100.066801} {\bibfield  {journal} {\bibinfo
  {journal} {Phys. Rev. Lett.}\ }\textbf {\bibinfo {volume} {100}},\ \bibinfo
  {pages} {066801} (\bibinfo {year} {2008})}\BibitemShut {NoStop}%
\bibitem [{\citenamefont {Vasenko}\ \emph {et~al.}(2015)\citenamefont
  {Vasenko}, \citenamefont {Basko},\ and\ \citenamefont
  {Hekking}}]{vasenko2015}%
  \BibitemOpen
  \bibfield  {author} {\bibinfo {author} {\bibfnamefont {A.~S.}\ \bibnamefont
  {Vasenko}}, \bibinfo {author} {\bibfnamefont {D.~M.}\ \bibnamefont {Basko}},
  \ and\ \bibinfo {author} {\bibfnamefont {F.~W.~J.}\ \bibnamefont {Hekking}},\
  }\href {\doibase 10.1103/PhysRevB.91.085310} {\bibfield  {journal} {\bibinfo
  {journal} {Phys. Rev. B}\ }\textbf {\bibinfo {volume} {91}},\ \bibinfo
  {pages} {085310} (\bibinfo {year} {2015})}\BibitemShut {NoStop}%
\bibitem{nonlin_note}%
  \BibitemOpen
  \bibinfo {note} {The non-linear thermovoltage though has been theoretically studied in discrete-level systems in Refs.~\cite{fazio2001,swirkowicz2009,kuo2010,sanchez2013,dutt2013,bjorn2013,Sanchez2014,zimbovskaya2016,erdman2017,daroca2018}.}\BibitemShut {Stop}%
\bibitem{fazio2001}
D. Boese, and R. Fazio,
\href{https://doi.org/10.1209/epl/i2001-00559-8}{Europhys. Lett. {\bf 56}, 576 (2001)}.
\bibitem{swirkowicz2009} 
R. \'Swirkowicz, M. Wierzbicki, and J. Barna\'s,
\href{http://dx.doi.org/10.1103/PhysRevB.80.195409}{Phys. Rev. B {\bf 80}, 195409 (2009)}.
\bibitem{kuo2010}
D. M.-T. Kuo, and Y.-C. Chang,
\href{https://doi.org/10.1103/PhysRevB.81.205321}{Phys. Rev. B {\bf 81}, 205321 (2010)}.
\bibitem{sanchez2013}
R. L\'{o}pez, and David S\'{a}nchez,
\href{http://doi.org/10.1103/PhysRevB.88.045129}{Phys. Rev. B {\bf 88}, 045129 (2013)}.
\bibitem{dutt2013}
P. Dutt, and K. Le Hur, 
\href{https://doi.org/10.1103/PhysRevB.88.235133}{Phys. Rev. B {\bf 88}, 235133 (2013)}.
\bibitem{bjorn2013}
R. S\'{a}nchez, B. Sothmann, A. N. Jordan, and M. B\"uttiker,
\href{https://doi.org/10.1088/1367-2630/15/12/125001}{New J. Phys. {\bf 15}, 125001 (2013)}.
\bibitem{Sanchez2014} 
M. A. Sierra and D. S\'{a}nchez,
\href{http://journals.aps.org/prb/abstract/10.1103/PhysRevB.90.115313}{Phys. Rev. B {\bf 90}, 115313 (2014)}.
\bibitem{zimbovskaya2016}
N. A. Zimbovskaya,
\href{http://dx.doi.org/10.1063/1.4922907 }{J. Chem. Phys. {\bf 142}, 244310 (2016)}.
\bibitem{erdman2017}
P. A. Erdman, F. Mazza, R. Bosisio, G. Benenti, R. Fazio, and F. Taddei,
\href{https://doi.org/10.1103/PhysRevB.95.245432}{Phys. Rev. B {\bf 95}, 245432 (2017)}.
\bibitem{daroca2018}
D. P. Daroca, P. Roura-Bas, and A. A. Aligia,
\href{https://doi.org/10.1103/PhysRevB.97.165433}{Phys. Rev. B {\bf 97}, 165433 (2018)}.
\bibitem{giazotto2006}
	F. Giazotto, T. T. Heikkil{\"a}, A. Luukanen, A. M. Savin, and J. P. Pekola, 
	\href{https://doi.org/10.1103/RevModPhys.78.217}{Rev. Mod. Phys. \textbf{78}, 217 (2006)}.
\bibitem{averin1991}
	D. V. Averin and K. K. Likharev, 
	\underline{Mesoscopic Phenomena in Solids} (North-Holland, Amsterdam, 1991).  
\bibitem [{\citenamefont {Nazarov}\ and\ \citenamefont
  {Banter}(2009)}]{nazarov2009}%
  \BibitemOpen
  \bibfield  {author} {\bibinfo {author} {\bibfnamefont {Y.~V.}\ \bibnamefont
  {Nazarov}}\ and\ \bibinfo {author} {\bibfnamefont {Y.~M.}\ \bibnamefont
  {Banter}},\ }\href@noop {} {\emph {\bibinfo {title} {Quantum Transport}}}\
  (\bibinfo  {publisher} {Cambridge, New York},\ \bibinfo {year}
  {2009})\BibitemShut {NoStop}%
\bibitem [{Note2()}]{suppl}%
  \BibitemOpen
  \bibinfo {note} {See Supplemental Material for details on the
  calculation of the charge and heat currents.}\BibitemShut {Stop}%
\bibitem [{Note1()}]{Note1}%
  \BibitemOpen
  \bibinfo {note} {The process where the same electron tunnels from $\alpha
  =L,R$ to I and then to $\beta =L,R$ is known as ``elastic co-tunneling'',
  since also the microscopic state on the island remains unchanged. As
  discussed in Refs.~\cite{nazarov2009,vasenko2015}, this
  process is relevant only when the thermal energy $k_BT$ and the voltage bias
  are much smaller than $\protect \sqrt {E_C\delta }$, where $\delta$ is the
  energy level spacing in the island. In our case the energy levels almost form
  a continuum, making $\protect \sqrt {E_C\delta }$ the smallest energy scale
  at play.}\BibitemShut {Stop}%
\bibitem [{\citenamefont {Kaasbjerg}\ and\ \citenamefont
  {Jauho}(2016)}]{kaasbjerg2016}%
  \BibitemOpen
  \bibfield  {author} {\bibinfo {author} {\bibfnamefont {K.}~\bibnamefont
  {Kaasbjerg}}\ and\ \bibinfo {author} {\bibfnamefont {A.-P.}\ \bibnamefont
  {Jauho}},\ }\href {\doibase 10.1103/PhysRevLett.116.196801} {\bibfield
  {journal} {\bibinfo  {journal} {Phys. Rev. Lett.}\ }\textbf {\bibinfo
  {volume} {116}},\ \bibinfo {pages} {196801} (\bibinfo {year}
  {2016})}\BibitemShut {NoStop}%
\bibitem [{\citenamefont {Bhandari}\ \emph {et~al.}(2018)\citenamefont
  {Bhandari}, \citenamefont {Chiriac\`o}, \citenamefont {Erdman}, \citenamefont
  {Fazio},\ and\ \citenamefont {Taddei}}]{bhandari2018}%
  \BibitemOpen
  \bibfield  {author} {\bibinfo {author} {\bibfnamefont {B.}~\bibnamefont
  {Bhandari}}, \bibinfo {author} {\bibfnamefont {G.}~\bibnamefont
  {Chiriac\`o}}, \bibinfo {author} {\bibfnamefont {P.~A.}\ \bibnamefont
  {Erdman}}, \bibinfo {author} {\bibfnamefont {R.}~\bibnamefont {Fazio}}, \
  and\ \bibinfo {author} {\bibfnamefont {F.}~\bibnamefont {Taddei}},\ }\href
  {\doibase 10.1103/PhysRevB.98.035415} {\bibfield  {journal} {\bibinfo
  {journal} {Phys. Rev. B}\ }\textbf {\bibinfo {volume} {98}},\ \bibinfo
  {pages} {035415} (\bibinfo {year} {2018})}\BibitemShut {NoStop}%
\bibitem [{\citenamefont {Bruus}\ and\ \citenamefont
  {Flensberg}(2004)}]{bruus2004}%
  \BibitemOpen
  \bibfield  {author} {\bibinfo {author} {\bibfnamefont {H.}~\bibnamefont
  {Bruus}}\ and\ \bibinfo {author} {\bibfnamefont {K.}~\bibnamefont
  {Flensberg}},\ }\href@noop {} {\emph {\bibinfo {title} {Many-body Quantum
  Theory in Condensed Matter Physics}}}\ (\bibinfo  {publisher} {Oxford
  University Press, New York},\ \bibinfo {year} {2004})\BibitemShut {NoStop}%
\end{thebibliography}

%

\clearpage
\begin{appendix}
\begin{widetext}

\section{Supplemental Material: Computing Charge and Heat Currents}
\label{app:rates}
The system is described by the following Hamiltonian
\begin{equation}
	\hat{H} = \sum_{\alpha=L,R}\hat{H}_{\alpha} + \hat{H}_\text{I} + \hat{H}_\text{t},
\end{equation}
where $\hat{H}_{\alpha} = \sum_{k\sigma} (\epsilon_k+eV_\alpha) a^{\dagger}_{k\sigma\alpha}a_{k\sigma\alpha}$ is the Hamiltonian of the free electrons in lead $\alpha=\text{L,R}$, $\hat{H}_\text{I} = \sum_{k\sigma} \epsilon_k c^\dagger_{k\sigma}c_{k\sigma} + E_C(\hat{n}-n_g)^2$ is the Hamiltonian of the electrons in the metallic island and $\hat{H}_\text{t} = \sum_{kp\sigma\alpha} t_{kp}^{(\alpha)} c^\dagger_{p\sigma}a_{k\sigma\alpha} + \text{h.c.}$ is the usual tunneling Hamiltonian between the leads and the island. $a_{k\sigma\alpha}$ ($a_{k\sigma\alpha}^\dagger$) is the destruction (creation) operator of electrons in lead $\alpha$ with energy $\epsilon_k + eV_\alpha$ and spin $\sigma$, $c_{k\sigma}$ ($c_{k\sigma}^\dagger$)  is the destruction (creation) operator of electrons in the metallic island with energy $\epsilon_k$ and spin $\sigma$, and $\hat{n}$ is the operator for the number of excess electrons on the island. 

In order to describe charge and heat transport in the system, we employ a master equation approach to compute the probabilities $P(n)$ in terms of all processes that can induce transitions between charges states (the tunneling rates). Sequential tunneling of electrons between the island and the leads changes the charge state by one, so it enters the master equation. Co-tunneling processes instead transfer an electron from one lead to another one via a virtual state in the island, but the overall process does not change the number of electrons in the island; consequently, the master equation does not depend on co-tunneling. Second order processes that transfer two electrons from/to the leads to/from the island can be safely neglected as the charging energy $E_C$ is much larger than the thermal energy $k_BT$ and than the voltage bias range considered in this work. The master equation reads 
\begin{equation}
	\frac{\partial P(n)}{\partial t} = \sum_{\alpha = L,R} \left\{ -P(n)\left[\Gamma_{\alpha I}(n) + \Gamma_{I \alpha}(n)\right]  + P(n-1)\Gamma_{\alpha I}(n-1) + P(n+1)\Gamma_{I \alpha}(n+1)  \right\},
	\label{eq:master}
\end{equation}
and we solve it by setting $\partial P(n)/\partial t=0$ for every $n$.  Eq.~(\ref{eq:master}) states that the probability of being in charge state $n$ can decrease (first r.h.s. term) if the island has $n$ excess charge states and an electron tunnels into or out of the island, while it can increase (second and third r.h.s. terms) if, after a sequential tunneling process, the number of excess charges on the island is $n$.  

Given the probabilities, the charge current can be computed using Eqs.~(\ref{eq:i_seq}) and (\ref{eq:i_cot}). The energy entering the metallic island $\dot{Q}_{\text{tun}}$ can be computed as
\begin{equation}
	\dot{Q}_{\text{tun}} \equiv I^\text{E}_\text{L} + I^\text{E}_\text{R} =  I^\text{h}_\text{L} + I^\text{h}_\text{R} + e(V_\text{L}-V_\text{R})I,
	\label{eq:qtun_energy}
\end{equation}
where $I^\text{E}_\alpha$  and $I^h_\alpha$ are respectively the energy (measured respect to the common voltage ground) and heat currents leaving reservoir $\alpha$, and we used the fact that $I^\text{E}_\text{L} = I^\text{h}_\text{L} + eV_\text{L}I$ and $I^\text{E}_\text{L} = I^\text{h}_\text{R} - eV_\text{R}I$.
We can simply interpret the r.h.s. of Eq.~(\ref{eq:qtun_energy}) by noticing that the heat entering the metallic island is given by the sum of the heat leaving the leads and the heat generated by Joule effect. We notice that a shift of the energy reference shifts $V_\text{L}$ and $V_\text{R}$, but it does not change $I^\text{h}_\text{L}$ and $I^\text{h}_\text{R}$, so $\dot{Q}_{\text{tun}}$, as defined in Eq.~(\ref{eq:qtun_energy}), does not depend on the un-physical energy reference.

The heat currents can be calculated in terms of ``heat rates''. We thus define $\Gamma^h_{\alpha \text{I}}(n)$ as the rate of heat leaving reservoir $\alpha$ when electrons tunnel sequentially from lead $\alpha$ to the island with $n$ initial electrons, and $\Gamma^h_{\text{I}\alpha}(n)$ as the rate of heat entering lead $\alpha$ when electrons tunnel sequentially from the island to lead $\alpha$ with $n$ initial electrons. Analogously, we define $\gamma^{h/\text{out}}_{\alpha\beta}(n)$ as the rate of heat leaving lead $\alpha$ when a co-tunneling process transfers one electron from lead $\alpha$ to lead $\beta$ with $n$ electrons in the island, and $\gamma^{h/\text{in}}_{\alpha\beta}(n)$ as the rate of heat entering lead $\beta$ when a co-tunneling process transfers one electron from lead $\alpha$ to lead $\beta$ with $n$ electrons in the island. Notice that also co-tunneling processes where $\alpha=\beta$ must be considered in the heat currents, since the electron leaving and the one entering the same lead can have different energies. Also the heat currents can be written as $I^h_\alpha = I^{h/\text{seq}}_\alpha+I^{h/\text{cot}}_\alpha$, where
\begin{equation}
	I^{h/\text{seq}} = \sum_n P(n)\left[\Gamma^h_{\alpha\text{I}}(n) - \Gamma^h_{\text{I}\alpha}(n)  \right]
\end{equation}
is the sequential-tunneling contribution, given by electrons tunneling between lead $\alpha$ and I, and
\begin{equation}
	I^{h/\text{cot}}_\alpha = \sum_{n,\beta=\text{L,R}} P(n)\left[\gamma_{\alpha \beta}^{h/\text{out}}(n) - \gamma_{\beta\alpha}^{h/\text{in}}(n)  \right]
\end{equation}
is the inelastic co-tunneling contribution. 

Using the $T$ matrix theory (or generalized Fermi golden rule) \cite{bruus2004,kaasbjerg2016,bhandari2018}, we can compute sequential and co-tunneling rates. The transition rate from a given initial state $\ket{i}$ to a final state $\ket{f}$ is given by
\begin{equation}
	\Gamma_{i\to f} = \frac{2\pi}{\hbar}p_i(1-p_f)\left|\braket{f|T|i}\right|^2 \delta(E_f-E_i),
	\label{eq:gen_rate}
\end{equation}
where $p_i$ and $p_f$ are the probabilities of finding the system in state $i$ and $f$, $E_i$ and $E_f$ are the energies of states $i$ and $f$ and $T = \hat{H}_t + \hat{H}_tG_0\hat{H}_t + \dots$ is the $T$ matrix with $G_0 = 1/(E_i-\hat{H}_0+i\eta)$ denoting the Green function in the absence of the $\hat{H}_t$, i.e. $\hat{H}_0 = \hat{H}_\text{L}+\hat{H}_\text{R} + \hat{H}_\text{I}$. We compute sequential rates by taking $T$ at first order in $\hat{H}_t$. We thus take $T=\hat{H}_t$ in Eq.~(\ref{eq:gen_rate}) and sum over all states in the lead and in the island, yielding
\begin{equation}
	\Gamma_{\alpha I}(n) = \frac{2\pi}{\hbar} \sum_{k_1\sigma_1,k_2\sigma_2} f_\alpha(\epsilon_{k_1})f^-_I(\epsilon_{k_2})\left|\braket{0|c_{k_2\sigma_2} H_t a^\dagger_{k_1\sigma_1\alpha}|0}\right|^2 \delta\left[\epsilon_{k_2}-\epsilon_{k_1} +\Delta E_\alpha(n)\right],
\end{equation}
where $\Delta E_\alpha(n) = U(n+1)-U(n) - eV_\alpha$ is the electrostatic energy difference to move an electron from lead $\alpha$ to the island, $f_{\alpha/\text{I}}(\epsilon) = [1+\exp{(\epsilon/(k_BT_{\alpha/I}))}]^{-1}$ is the Fermi distribution of lead $\alpha$ at temperature $T_\alpha$ or of the island at temperature $T_\text{I}$, $f^-_{\alpha/\text{I}}(\epsilon) =f_{\alpha/\text{I}}(-\epsilon)= 1-f_{\alpha/\text{I}}(\epsilon)$, and $k_B$ is the Boltzmann constant. 
An analogous expression holds for $\Gamma_{I \alpha}(n)$. The heat rates are computed in the same way, taking into account that an amount of heat $\epsilon_k$ is removed(injected) from(into) a lead if an electron with momentum $k$ tunnels from(into) the lead. We thus have that
\begin{equation}
\begin{aligned}
	\Gamma^h_{\alpha I}(n) &= \frac{2\pi}{\hbar} \sum_{k_1\sigma_1,k_2\sigma_2} \epsilon_{k_1} f_\alpha(\epsilon_{k_1})f^-_I(\epsilon_{k_2})\left|\braket{0|c_{k_2\sigma_2} H_t a^\dagger_{k_1\sigma_1\alpha}|0}\right|^2 \delta\left[\epsilon_{k_2}-\epsilon_{k_1} +\Delta E_\alpha(n)\right], \\
	\Gamma^h_{I \alpha}(n) &= \frac{2\pi}{\hbar} \sum_{k_1\sigma_1,k_2\sigma_2} \epsilon_{k_2} f_\text{I}(\epsilon_{k_1})f^-_\alpha(\epsilon_{k_2})\left|\braket{0|a_{k_2\sigma_2\alpha} H_t c^\dagger_{k_1\sigma_1} |0}\right|^2 \delta\left[\epsilon_{k_2}-\epsilon_{k_1} -\Delta E_\alpha(n-1)\right].
\end{aligned}
\end{equation} 
By assuming that the energy levels in the leads and in the island form a continuum, by taking a constant density of states around the Fermi energy and by replacing the hopping parameters $t_{kp}^{(\alpha)}$ with their averaged value over $k$ and $p$, we can write the sequential rates and heat rates in terms of the functions
\begin{equation}
\begin{aligned}
	\Upsilon_\alpha(\Delta E) \equiv \frac{1}{e^2R_\alpha} \int\limits_{-\infty}^{+\infty} d\epsilon f_\alpha(\epsilon) f^-_\text{I}(\epsilon-\Delta E), \\
	\Upsilon^h_\alpha(\Delta E) \equiv \frac{1}{e^2R_\alpha} \int\limits_{-\infty}^{+\infty} d\epsilon \,\epsilon f_\alpha(\epsilon) f^-_\text{I}(\epsilon-\Delta E), \\
\end{aligned}
	\label{eq:rate_se}
\end{equation}
where $R_\alpha$ is the tunnel resistance between lead $\alpha$ and the island, as follows:
\begin{equation}
\begin{aligned}
	&\Gamma_{\alpha\text{I}}(n) = \Upsilon_\alpha[\Delta E_\alpha(n)], &\Gamma_{\text{I}\alpha}(n+1) = \Upsilon_\alpha[-\Delta E_\alpha(n)], \\
	&\Gamma^h_{\alpha\text{I}}(n) = \Upsilon^h_\alpha[\Delta E_\alpha(n)], &\Gamma^h_{\text{I}\alpha}(n+1) = -\Upsilon^h_\alpha[-\Delta E_\alpha(n)].
\end{aligned}
\end{equation}

Co-tunneling rates are second order processes that involve initial and final states with two electrons, so we now consider $T = \hat{H}_tG_0\hat{H}_t$. We thus take $\ket{i} = a^\dagger_{k_1\sigma_1\alpha}c^\dagger_{q_1\tau_1}\ket{0}$ and $\ket{f} = a^\dagger_{q_2\tau_2\beta}c^\dagger_{k_2\sigma_2}\ket{0}$, which corresponds to considering the process where an electron in state $k_1\sigma_1$ tunnels from lead $\alpha$ to the island into state $k_2\sigma_2$, and another one coherently tunnels from the island in state $q_1\tau_1$ to lead $\beta$ into state $q_2\tau_2$. From Eq.~(\ref{eq:gen_rate}) we have that
\begin{equation}
	\gamma_{\alpha\beta}(n) =  \frac{2\pi}{\hbar} \sum_{\substack{k_1\sigma_1,k_2\sigma_2 \\ q_1\tau_1,q_2\tau_2}} f_\alpha(\epsilon_{k_1})f_\text{I}(\epsilon_{q_1})f^-_\beta(\epsilon_{q_2})f^-_\text{I}(\epsilon_{k_2})
	\left| \sum_{\nu}  \frac{\braket{f |H_t |\nu}\braket{\nu|H_t |i}}{E_i-E_\nu+i\eta}   \right|^2 \delta\left[\epsilon_{q_2}+\epsilon_{k_2}-\epsilon_{q_1}-\epsilon_{k_1} +e(V_\beta-V_\alpha)\right],
	\label{eq:cot_rate}
\end{equation}
where $E_i= \epsilon_{q_1} + \epsilon_{k_1} + eV_\alpha + U(n)$ is the energy of state $\ket{i}$, the sum over $\ket{\nu}$ runs over a complete set of eigenstates $\{ \ket{\nu} \}$ of $H_0$, and $E_\nu$ is the energy, evaluated with $H_0$, of state $\ket{\nu}$. As we did for the sequential rates, we notice that in the processes described in Eq.~(\ref{eq:cot_rate}), the heat leaving reservoir $\alpha$ is $\epsilon_{k_1}$, while the heat entering reservoir $\beta$ is $\epsilon_{q_2}$. The co-tunneling heat rate leaving reservoir $\alpha$, $\gamma_{\alpha\beta}^{h/\text{out}}(n)$, is thus given by Eq.~(\ref{eq:cot_rate}) adding an $\epsilon_{k_1}$ inside the sum over the initial and final states, while the co-tunneling heat rate entering reservoir $\beta$, $\gamma_{\alpha\beta}^{h/\text{in}}(n)$, is also given by Eq.~(\ref{eq:cot_rate}) adding an $\epsilon_{q_2}$ inside the sum over the initial and final states. Manipulating Eq.~(\ref{eq:cot_rate}) using the same approximations mentioned for the sequential rates, we find that by defining
\begin{equation}
\begin{aligned}
	\upsilon_{\alpha\beta}(\Delta E,\Delta E_1,\Delta E_2) &= \frac{\hbar}{2\pi} \int\limits_{-\infty}^{+\infty} d\epsilon \Upsilon_\alpha(-\epsilon)\Upsilon_\beta(\epsilon+\Delta E)  \left| \frac{1}{\epsilon+\Delta E_1-i\eta} - \frac{1}{\epsilon-\Delta E_2+\Delta E+i\eta} \right|^2,  \\
	\upsilon^{h/\text{out}}_{\alpha\beta}(\Delta E,\Delta E_1,\Delta E_2) &= \frac{\hbar}{2\pi} \int\limits_{-\infty}^{+\infty} d\epsilon \Upsilon^h_\alpha(-\epsilon)\Upsilon_\beta(\epsilon+\Delta E)  \left| \frac{1}{\epsilon+\Delta E_1-i\eta} - \frac{1}{\epsilon-\Delta E_2+\Delta E+i\eta} \right|^2,  \\
	\upsilon^{h/\text{in}}_{\alpha\beta}(\Delta E,\Delta E_1,\Delta E_2) &= -\frac{\hbar}{2\pi} \int\limits_{-\infty}^{+\infty} d\epsilon \Upsilon_\alpha(-\epsilon)\Upsilon^h_\beta(\epsilon+\Delta E)  \left| \frac{1}{\epsilon+\Delta E_1-i\eta} - \frac{1}{\epsilon-\Delta E_2+\Delta E+i\eta} \right|^2, 	
 \end{aligned}
 \label{eq:cot_integrals}
\end{equation}
we can write the co-tunneling rates and heat rates as
\begin{align}
	\gamma_{\alpha\beta}(n) &= \upsilon_{\alpha\beta}\left[e(V_\beta-V_\alpha), \Delta U_\alpha(n), -\Delta U_\beta(n-1)   \right], \\
	\gamma_{\alpha\beta}^{h/\text{out}}(n) &= \upsilon^{h/\text{out}}_{\alpha\beta}\left[e(V_\beta-V_\alpha), \Delta U_\alpha(n), -\Delta U_\beta(n-1)   \right], \\
		\gamma_{\alpha\beta}^{h/\text{in}}(n) &= \upsilon^{h/\text{in}}_{\alpha\beta}\left[e(V_\beta-V_\alpha), \Delta U_\alpha(n), -\Delta U_\beta(n-1)   \right].
\end{align}
At last, we notice that the integrals in Eq.~(\ref{eq:cot_integrals}) are divergent in the limit $\eta\to 0^+$. In order to overcome this problem, we adopt a commonly used ``regularization scheme'' \cite{turek2002,koch2004,nazarov2009,kaasbjerg2016,bhandari2018}. All three integrals can be written in the form
\begin{multline}
	\mathcal{I} = \int\limits_{-\infty}^{+\infty} d\epsilon\, g(\epsilon) \left| \frac{1}{\epsilon-\alpha_1-i\eta} - \frac{1}{\epsilon-\alpha_2+i\eta} \right|^2  = \\
	 \sum_{i=1,2} \left\{ \int\limits_{-\infty}^{+\infty} d\epsilon\, g(\epsilon) \left| \frac{1}{\epsilon-\alpha_i-i\eta} \right|^2\right\} 
	 - 2\int\limits_{-\infty}^{+\infty} d\epsilon\, g(\epsilon) \Re{\left\{ \frac{1}{(\epsilon-\alpha_1+i\eta)(\epsilon-\alpha_2+i\eta)} \right\}} =  \sum_{i=1,2} \left\{ \mathcal{I}^{(1)}_i  \right\} -2 \mathcal{I}^{(2)}
\end{multline}
where $g(\epsilon)$ is a suitable function and $\alpha_1$ and $\alpha_2$ are suitable constants. We now analyze each integral:
\begin{multline}
	\mathcal{I}^{(1)}_i = \int\limits_{-\infty}^{+\infty} d\epsilon\, g(\epsilon) \left| \frac{1}{\epsilon-\alpha_i-i\eta} \right|^2 
	= \int\limits_{-\infty}^{+\infty} d\epsilon \frac{g(\epsilon) - g(\alpha_i) + g(\alpha_i)}{(\epsilon-\alpha_i)^2+\eta^2} 
	= g(\alpha_i)\int\limits_{-\infty}^{+\infty} d\epsilon \frac{1}{(\epsilon-\alpha_i)^2+\eta^2}   +   \int\limits_{-\infty}^{+\infty} d\epsilon \frac{g(\epsilon) - g(\alpha_i)}{(\epsilon-\alpha_i)^2+\eta^2}  \\
	= \frac{\pi g(\alpha_i)}{\eta} + \mathcal{P} \int\limits_{-\infty}^{+\infty} d\epsilon \frac{g(\epsilon) - g(\alpha_i)}{(\epsilon-\alpha_i)^2} + O(\eta),
	\label{eq:regul}
\end{multline}
where $\mathcal{P}$ denotes a principal value integration. We notice that the last step of Eq.~(\ref{eq:regul}) is an expansion for small $\eta$. In particular, the first term diverges as $1/\eta$, the second one is finite and independent of $\eta$, while the third one goes to zero if $\eta\to 0$. The regularization scheme consists of dropping the divergent term and retaining only the second term, which is finite and independent of $\eta$:
\begin{equation}
	\mathcal{I}^{(1)}_i \to  \mathcal{P} \int\limits_{-\infty}^{+\infty} d\epsilon \frac{g(\epsilon) - g(\alpha_i)}{(\epsilon-\alpha_i)^2}.
\end{equation}
 Let's now turn to
\begin{multline}
	\mathcal{I}^{(2)} = \int\limits_{-\infty}^{+\infty} d\epsilon\, g(\epsilon) \Re{\left\{ \frac{1}{(\epsilon-\alpha_1+i\eta)(\epsilon-\alpha_2+i\eta)} \right\}} =
	\int\limits_{-\infty}^{+\infty} d\epsilon  g(\epsilon) \frac{ (\epsilon-\alpha_1)(\epsilon-\alpha_2)-\eta^2 }{\left[(\epsilon-\alpha_1)(\epsilon-\alpha_2) -\eta^2 \right]^2 + \eta^2\left[(\epsilon-\alpha_1) +(\epsilon-\alpha_2) \right]^2} = \\
	 \int\limits_{-\infty}^{+\infty} d\epsilon g(\epsilon)  \frac{(\epsilon-\alpha_1)(\epsilon-\alpha_2)-\eta^2  }{(\epsilon-\alpha_1)^2(\epsilon-\alpha_2)^2 + \eta^2\left[(\epsilon-\alpha_1)^2 +(\epsilon-\alpha_2)^2 + \eta^2 \right]}.
	 \label{eq:regul_2}
\end{multline}
We notice that the denominator in Eq.~(\ref{eq:regul_2}) is always positive and non-zero. In the limit $\eta\to 0$, the term proportional to $(\epsilon-\alpha_1)(\epsilon-\alpha_2)$ turns into a principal value integration, while the term proportional to $-\eta^2$ vanishes. The regularization scheme thus consists of 
\begin{equation}
	\mathcal{I}^{(2)} \to \mathcal{P}\int\limits_{-\infty}^{+\infty} d\epsilon  \frac{ g(\epsilon)}{(\epsilon-\alpha_1)(\epsilon-\alpha_2)},
\end{equation}
which is now finite and independent of $\eta$.

\end{widetext}
\end{appendix}

\end{document}